\documentclass[aps,prd,onecolumn,preprintnumbers,eqsecnum,nofootinbib,showpacs]{revtex4-1}
 
\usepackage[reqno]{amsmath}
\usepackage{ulem}
\usepackage{amssymb}
\usepackage{graphicx}
\usepackage{color}
\usepackage[usenames,dvipsnames]{xcolor}
\usepackage[colorlinks=true,urlcolor=RoyalBlue,citecolor=RoyalBlue,linkcolor=black,linktocpage=true]{hyperref}

\newcommand{\arxiv}[1]{\href{http://arxiv.org/abs/#1}{\textcolor{RoyalBlue}{arXiv:#1}}}

\newcommand{\HEParxiv}[1]{\href{http://arxiv.org/abs/hep-ph/#1}{\textcolor{RoyalBlue}{hep-ph/#1}}}
\newcommand{\EXarxiv}[1]{\href{http://arxiv.org/abs/hep-ex/#1}{\textcolor{RoyalBlue}{hep-ex/#1}}}

\newcommand{\MeV}{\,{\rm MeV}}
\newcommand{\GeV}{\,{\rm GeV}}
\newcommand{\TeV}{\,{\rm TeV}}

\newcommand{\beq}{\begin{equation}}
\newcommand{\eeq}{\end{equation}}

\newcommand{\bw}{\begin{widetext}}
\newcommand{\ew}{\end{widetext}}

\newcommand{\bea}{\begin{eqnarray}}
\newcommand{\eea}{\end{eqnarray}}
\newcommand{\ba}{\begin{array}}
\newcommand{\ea}{\end{array}}

\newcommand{\ovl}{\overline}
\newcommand{\eps}{\epsilon}
\newcommand{\al}{\alpha}
\newcommand{\be}{\beta}
\newcommand{\lam}{\lambda}

\newcommand{\im}{{\rm Im}\,}
\newcommand{\re}{{\rm Re}\,}

\newcommand{\Fs}{\mathcal{F}}
\newcommand{\mup}{\mu^{\prime}}
\newcommand{\mupp}{\mu^{\prime\prime}}

\newcommand{\dmsol}{\mbox{$\Delta m^2_{\odot}$}}
\newcommand{\dma}{ \mbox{$\vert\Delta m^2_{\rm A}\vert$}}

\begin{document}
\preprint{CFTP/12-016}
\preprint{TUM-HEP 862/12}
\title{Triplet Scalar Dark Matter and Leptogenesis in an  Inverse See-Saw Model of Neutrino Mass Generation}
\author{Fran\c cois-Xavier~Josse-Michaux}
\email{fxjossemichaux@gmail.com}
\affiliation{Centro de F\'isica Te\'orica de Part\'iculas CFTP, Instituto Superior Tecnico, Universidade Tecnica de Lisboa, 1049-001 Lisbon, Portugal}
\author{Emiliano~Molinaro}
\email{emiliano.molinaro@tum.de}
\affiliation{Physik-Department T30d, Technische Universit\"at M\"unchen, James-Franck Stra\ss e 1, 85748 Garching, Germany}
\begin{abstract}
We propose a UV completion of the inverse see-saw scenario using fermion SU(2)$_{\rm L}$ triplet representations. 
 Within this framework, a \textit{variation} of the standard thermal leptogenesis is achievable at the $\mathcal{O}$(TeV) scale, owing to the presence of a viable dark matter candidate. This baryogenesis scenario is ruled out if a triplet fermion is observed at the LHC.
The dark matter is given by the lightest neutral component of a complex scalar SU(2)$_{\rm L}$ triplet, with mass $m_{\rm DM}$ in the TeV range.
The scalar sector, which is enriched in order to account for the small neutrino masses, is treated in detail and shows potentially 
 sizable Higgs boson $h \to \gamma \gamma$ rates together with large  $h$ invisible branching ratios.
\end{abstract}


\maketitle

\section{Introduction}

The recent discovery of a new scalar resonance by ATLAS~\cite{ATLAS} and CMS~\cite{CMS} experiments at the Large Hadron Collider (LHC) 
may finally allow us to disclose the mechanism of electroweak symmetry breaking (EWSB) and fully test the Standard Model (SM) scalar sector.
On the other hand, new physics beyond the SM must be advocated if we want to explain neutrino masses and mixing, dark matter (DM) and the
baryon asymmetry of the universe (BAU).

In this article, we propose a simple renormalizable extension of the SM where all these issues are addressed:
\begin{itemize}
	\item[1.] The small neutrino masses are explained by an inverse see-saw model, constructed by the addition of fermion and scalar fields to the SM particle content.
	\item[2.] The DM corresponds to the neutral component of a complex triplet scalar field, 
	explaining the observed relic abundance: $\Omega_{\rm DM}h^2 =0.112\pm 0.006$~\cite{PDG12}.
	\item[3.] The BAU $\Omega_{\rm B} h^{2} = 0.0226\pm 0.006$~\cite{PDG12} is explained through a  \textit{variation}  of the standard leptogenesis mechanism we proposed in~\cite{Previous}.
\end{itemize}
The model Lagrangian is invariant under a global U(1)$_{\rm X}$ symmetry, which is spontaneously broken at the electroweak scale to a remnant $Z_{2}$ symmetry.
The role of the additional U(1)$_{\rm X}$ is twofold: through its breaking, it ensures the stability of the triplet DM candidate and provides tiny light neutrino masses. With the new fields typically at the TeV scale, the model actually provides a  UV completion of the inverse see-saw scenario~\cite{Mohapatra:1986bd}. We list in Table~\ref{FieldAssign} the fermion and scalar content of our model and their charges under SU$(2)_{\rm L}\times$U$(1)_{\rm Y}\times[$U$(1)_{\rm X}]$. It is clear from this table that the U$(1)_{\rm X}$ charge effectively
corresponds to a generalization of the baryon-lepton charge $B-L$. 

The heavy right-handed (RH) neutrinos $N_{1,2}$ we introduce in Table I are charged under U$(1)_{\rm X}$, which is conserved above the EWSB.
Therefore, thermal leptogenesis~\cite{lepto,Davidson:2008bu} cannot be realized as in the standard type I~\cite{SeesawI} or type III~\cite{SeesawIII} see-saw extensions of the SM.
Similarly to scenarios where the $B-L$ charge is conserved~\cite{DiracLepto1,DiracLepto2,DiracLepto3,Sahu:2008aw,GonzalezGarcia:2009qd, Kohri:2009yn}, a nonzero lepton number asymmetry can be generated by out of equilibrium decays of a heavy scalar/fermion particle, while a (model-dependent)  mechanism prevents the washout of the total lepton charge~\footnote{In the case of Dirac leptogenesis, very small neutrino Yukawa couplings prevent equilibration between left and right-handed lepton asymmetries, until long after EWSB when sphalerons are no longer active. A similar role for the sphaleron decoupling epoch is used in~\cite{GonzalezGarcia:2009qd}, in interplay with lepton  flavour effects~\cite{LeptonFla1,LeptonFla2}.}.
The generated lepton asymmetry is converted into a nonzero baryon number  by the rapid nonperturbative sphaleron processes~\cite{Sphaleron1,Sphaleron2}.

In the present case, we consider a variant of our work~\cite{Previous}, using$-$instead of singlets$-$SU$(2)_{\rm L}$ triplet representations for the RH neutrinos $N_{1,2}$ and the scalar $S$~\footnote{Triplet SU$(2)_{\rm L}$ fermion representations are used in the inverse see-saw mechanism, e.g. in~\cite{Ma:2009kh,Ibanez:2009du}}.
 A singlet Majorana fermion $N_3$ is introduced, and as in type I leptogenesis its decays produce asymmetries in $N$ and $S$.
 Successful leptogenesis is possible due to the transfer~\footnote{A similar $B-L$ conserving Yukawa-driven transfer mechanism for leptogenesis is studied in~\cite{Sahu:2008aw}, based on a radiative model of neutrino masses.} of the U$(1)_{\rm X}$ charge asymmetry of $N$ to the SM leptons, through (fast) neutrino Yukawa interactions.

The inverse see-saw completion we propose requires several additional scalar particles.
Such an enlarged scalar sector induces deviations from SM expectations, and after the observation of a $\sim 126$ GeV boson~\cite{ATLAS,CMS}, a reanalysis of the scalar spectrum of~\cite{Previous} is in due order.
 As we show below, a large Higgs diphoton rate is possible, together with a large Higgs invisible branching ratio. 

This paper is organized as follows: in Sec.~\ref{Sec2} we describe the scalar sector of the model and compare its predictions to current ATLAS and CMS data. In Sec.~\ref{Section:DM} we outline the phenomenology of the triplet scalar dark matter scenario.
The neutrino mass generation and  the leptogenesis mechanism are reported in Sec.~\ref{Sect3}.
Finally, we summarize the main results of this work in the concluding section.

\section{Scalar Sector}\label{Sec2}

We extend the Standard Model particle content with new scalar representations of the SM gauge group, which are listed in Table~\ref{FieldAssign}:
\begin{itemize}
\item[i)] a $\rm SU(2)_{\rm L}$  doublet $H_{2}^{\rm T}\equiv\left(H_{2}^{+}\;\,H_{2}^{0} \right)$, with hypercharge Y=1/2, in addition to the SM Higgs doublet 
$H_{1}^{\rm T}\equiv\left(H_{1}^{+}\;\,H_{1}^{0} \right)$,
\item[ii)] a complex singlet $\phi$, 
\item[iii)] a complex ${\rm SU(2)_{\rm L}}$  triplet $S$ with zero hypercharge.
\end{itemize}
All these new fields have a nonzero charge under a global ${\rm U(1)_{\rm X}}$ symmetry.
In this scenario, the presence of $H_{2}$ and $\phi$ is motivated by the requirement of generating light neutrino masses through 
the (inverse) see-saw mechanism and the possibility of realizing  successful leptogenesis (see Sec.~\ref{Sect3}). 
We list in Table~\ref{FieldAssign} the particle content of the model with their charge assignments under SU$(2)_{\rm L}$, U$(1)_{\rm Y}$ and the global U$(1)_{\rm X}$.
\begin{table}[!t]
\begin{center}
\begin{tabular}{|c||c|c|c|c|c||c|c|c||c|}
\hline 
\rule[0.15in]{0cm}{0cm}{\tt Field}& $\ell_{\alpha}$ & $e_{R\alpha}$ & $N_{1}$ & $N_{2}$& $N_{3}$ & $H_{1}$ & $H_{2}$ & 
$\phi$ & $S$  \\
\hline
$\rm SU(2)_{\rm L}$ & $\mathbf{2}$ & $\mathbf{1}$ & $\mathbf{3}$ & $\mathbf{3}$& $\mathbf{1}$ & $	\mathbf{2}$ & $\mathbf{2}$ & 
$\mathbf{1}$ & $\mathbf{3}$
\\\hline
$\rm U(1)_{\rm Y}$ & -$1/2$ & -$1$ & $0$ & $0$ & $0$& $1/2$ & $1/2$ & 
$0$ & $0$
\\\hline
$\rm U(1)_{\rm X}$ & -$1$ & -$1$ & -$1$  & +$1$& $0$ & $0$ & $2$ & 
-$2$ & -$1$
\\\hline
\end{tabular}
\caption{Particle content and their charge assignments.}\label{FieldAssign}
\end{center}
\end{table}%

The most general scalar potential which is invariant under the SU$(2)_{\rm L}\times$U$(1)_{\rm Y}\times$[U$(1)_{\rm X}$] is conveniently split into two parts:
$\mathcal{V}_{\rm SC}\equiv \mathcal{V}_{\rm SB}\,+\,\mathcal{V}_{\rm DM}$.
The scalar potential pertaining to DM, $\mathcal{V}_{\rm DM}$, is discussed in Sec.~\ref{Section:DM}. The symmetry breaking scalar potential $\mathcal{V}_{\rm SB}$ is~\cite{Previous}:
\bw
\bea
\mathcal{V}_{\rm SB}&=&-\mu_{1}^2\,(H_1^{\dag}\,H_1) + \lam_{1}\, (H_1^{\dag}\,H_1)^2 - \mu_{2}^2\,( H_2^{\dag}\,H_2) + \lam_{2}\, (H_2^{\dag}\,H_2)^2 - \mu_{3}^2\, \phi^{*} \phi + \lam_{3}\, (\phi^{*} \phi)^2\nonumber \\  &+& \kappa_{12}\, (H_1^{\dag}\,H_1) \,(H_2^{\dag}\,H_2) +\kappa_{12}^{\prime}\, (H_1^{\dag}\,H_2)\,(H_2^{\dag}\,H_1)+\kappa_{13}\, (H_1^{\dag}\,H_1) \phi^{*} \phi + \kappa_{23}\, (H_2^{\dag}\,H_2) \phi^{*} \phi \nonumber \\
&-&\frac{\mup}{\sqrt{2}}\, \left( (H_1^{\dag}\,H_2) \phi + (H_2^{\dag}\,H_1) \phi^{*}\right)\,.\label{VSB}
\eea
\ew
Only the scalar fields  with even ${\rm U(1)_{\rm X}}$ charge acquire a nonzero (real) vacuum expectation value (vev): the two SU$(2)_{\rm L}$ doublets $H_{1,2}$ and the singlet $\phi$~\cite{Previous}, their respective vev being $v_{1,2}$ and $v_{\phi}$.
With such a charge assignment, the global $\rm U(1)_{\rm X}$ is broken down to a $Z_{2}$ which stabilizes the DM, odd under ${\rm U(1)_{\rm X}}$.
Given the field content and the charge assignment reported in Table~\ref{FieldAssign}, this model can be thought as a variant of the type I two-Higgs doublet model (see e.g. \cite{Branco:2011iw}),
augmented with a further complex scalar $\phi$~\footnote{Such scalar spectrum is considered in a different context in~\cite{Carmi:2012in}.}: both doublets $H_{1,2}$ couple to gauge bosons and while only $H_1$ couples to SM fermions via Yukawa couplings, $H_2$ and $\phi$ couple to the new particles $S$, $N_D$ and $N_3$ (see the following sections).

We first discuss the scalar spectrum arising from the spontaneous breaking of  SU$(2)_{\rm L}\times$U$(1)_{\rm Y}\times$[U$(1)_{X}$] to $\rm U(1)_{\rm em}\times$[$Z_{2}$] and
we postpone the analysis of the triplet scalar $S$ to the next section.
The spectrum consists of \cite{Previous}:
\begin{itemize}
\item[-] 1 charged scalar $H^\pm$,
\item[-] 3 CP-even neutral scalars $h^0$, $H^0$ and $h_A$,
\item[-] 2 pseudoscalars $A^0$ and $J$.
\end{itemize} 
The latter is the Goldstone boson associated with the breaking of the global ${\rm U(1)_{\rm X}}$ symmetry and is usually dubbed Majoron in theories with spontaneously broken lepton charge~\cite{Majoron}.
Since it is a massless particle at the renormalizable level, strong constraints apply on its couplings to SM fermions:  a hierarchical pattern for the vevs of the scalar fields, $v_2 \ll v_{1}, v_{\phi}$, is required~\footnote{A suppressed value of $v_{2}$ is naturally 
realized from the minimization of the potential (\ref{VSB}), due to the symmetries of the model~\cite{Previous}.}.
In the limit of negligible $v_2$, the longitudinal gauge boson components are  $W_{L}^\pm\sim H_1^\pm$ and $Z_L \sim \sqrt{2}\,\im{(H_1^0)}$, while the scalar mass eigenstates are to a good approximation:
\bea
H^\pm \sim H_2^\pm\,,\quad   h_A \sim \sqrt{2}\;\re{(H_2^0})\,,\quad A_0 \sim \sqrt{2}\;\im{(H_2^0)}\,\quad {\rm and}\quad J \sim \sqrt{2}\; \im{(\phi)},\nonumber
\eea
while the two neutral scalars $h^{0}$ and $H^{0}$ are related to the interaction fields $H_{1}$ and $\phi$ by  a rotation
\bea
\left(
\begin{array}{c} h^0 \\ H^0 \end{array} \right)=R(-\theta)\,\left(\begin{array}{c} \sqrt{2}\,\re{(H_1^0}) \\ \sqrt{2}\,\re{(\phi)} \end{array} \right)\,,\label{rot13}
\eea
where $\theta$ is a function of the vevs $v_{1,\phi}$ and the quartic couplings of $H_{1}$ and $\phi$ in $\mathcal{V}_{\rm SB}$~\cite{Previous}.
Typically, we have $v_2 \lesssim 10$ MeV, $v_1\simeq 246.2$ GeV, while $v_\phi$ is free.
Recalling that only $H_1$ has Yukawa couplings to SM fermions (cf. Table~\ref{FieldAssign}), $h_A$, $A^0$ and $H^\pm$ couple to the SM sector only through gauge interactions and via the scalar quartic couplings, while $h^0$ and $H^0$ can have \textit{a priori} sizable Yukawa couplings to SM fermions.

\subsection{LHC constraints}

 In~\cite{Previous}  we performed a detailed analysis of the scalar spectrum and  we constrained the parameter space given the experimental results available at that time.
 However, after the observation of a new scalar particle with mass $m_{h} \sim 126 \GeV$  by both ATLAS~\cite{ATLAS} and CMS~\cite{CMS} Collaborations, it is worth studying more carefully the scalar mass spectrum of the model.

We assume the new discovered particle corresponds to the lightest of the two CP-even scalars with possibly large couplings to SM fermions, that is $h^0$.
 We ought to verify whether the model can explain ATLAS and CMS data,  while encompassing present phenomenological constraints.
In order to proceed with the analysis, we consider the ratios between the production of a boson $H$ decaying into a generic final SM state $i$  to  the corresponding SM prediction,
\bea\label{ratioI}
\mu_i(H) \equiv \frac{\sigma(pp\to H)_i\times {\rm Br}(H \to i)}{\sigma(pp\to h)_i^{\rm SM}\times {\rm Br}(h \to i)^{\rm SM}}\,.
\eea
\hspace{1mm}

Depending on the decay products, Higgs searches target specific production channels, hence the labeling 
$\sigma(pp\to h)_i$.
Notice that, in our model, the Higgs signal strengths $\mu_i$s may be affected in several aspects and may differ with respect to the SM predictions.

Regarding the Higgs boson production, as no extra colored particles are introduced, there are no new contributions to the loop-induced gluon-gluon fusion process. 
Furthermore, from the definition, eq.~(\ref{rot13}), the couplings of $h^0$ ($H^0$) to the SM particles are those of $H_1$ times $\cos(\theta)$  $(\sin(\theta))$: all production channels are thus equally rescaled and we have
\bea\label{ratioII}
\frac{\sigma(pp\to h^0 )_i}{\sigma(pp\to h)^{\rm SM}_i}= \cos^{2}(\theta)\,,\quad \frac{\sigma(pp\to H^0)_i}{\sigma(pp\to h)^{\rm SM}_i}= \sin^{2}(\theta)\,.\nonumber
\eea
The branching ratios ${\rm Br}(H \to i)$ reported in (\ref{ratioI}) are affected in three ways.
First, as $h^0~(H^0)$ couplings to SM particles are rescaled by $\cos{(\theta)}$ $(\sin{(\theta)} )$ compared to the SM ones, tree level Higgs decays to SM particles are rescaled by $\cos{(\theta)}^2$ $(\sin{(\theta)}^2 )$.
Second and \textit{ceteris paribus}, the branching ratios are reduced with respect to the corresponding SM predictions, ${\rm Br}(h \to i)^{\rm SM}$, because of the presence of new decay channels.
 For the sake of simplicity, we assume that $h^0$ is the lightest massive neutral scalar particle, thus closing these decay channels.
 Moreover, the invisible decays $h^{0}/H^{0}\to J \,J$ can be sizable in some regions of the parameter space.
Third, deviations from the SM occur in loop-induced processes.
As already stated, no extra colored particles are introduced, so  Higgs decays to gluons are not affected.
On the other hand, the diphoton decay channel $h^0~(H^0) \to \gamma \gamma$ is affected by the presence of new charged particles.
Several works have emphasized possible deviations from SM expectations of the diphoton decay rate, due to the presence of extra  fermion/scalar charged particles (see e.g. \cite{Giardino:2012ww,Espinosa:2012vu,Espinosa:2012im,diphoton1,diphoton8}).
 In our model, we potentially have to consider the effect of 5 extra charged particles: the scalar $H^\pm$ originating from the doublet $H_{1,2}$, the two scalars $S_L^\pm$ and $S_H^\pm$ coming from our DM scalar triplet (see Sec.~\ref{Section:DM}), and the two charged fermions $\Sigma_{1,2}^\pm$ coming from the triplet $N_D$ introduced in our type III inverse see-saw variant (see Sec.~\ref{Sect3}).
  As discussed in the next sections, the triplet particles have large masses, typically $\mathcal{O}(\TeV)$: their contribution to the  diphoton decay rate is therefore negligible.
Oppositely~-~and as in the case of  type I  2HDM~\footnote{Flavour physics observables in our case do not put constraints on $H^\pm$ mass, given the large vev hierarchy: $v_2 \ll v_1$.}-~the charged scalar $H^{\pm}$  can be sufficiently light to produce observable effects in the  $h^{0}\to \gamma\gamma$ decay rate~\footnote{LHC constraints on $m_{H^\pm}$ are yet based on the decays $t\to H^+ b$: the values we obtain are well below the present bound: Br$(t\to H^+ b) \lesssim 10^{-2}$~\cite{LHChp1,LHChp2}.}: assuming ${\rm Br}(H^+ \to  c\,\ovl{s})+{\rm Br}(H^+ \to \tau^+\,\nu )=1$, LEP2 derived a conservative lower bound on the mass of $H^{\pm}$, $m_{H^\pm} \gtrsim 80 \GeV$ \cite{Searches:2001ac}. 
The Higgs diphoton rate is given in our model by \cite{Djouadi,Barroso:2012wz}~\footnote{We use the same conventions and notations as~\cite{Barroso:2012wz}.}
\bw
\bea
\Gamma(h^0/H^0\to \gamma \gamma)&=&\frac{G_\mu\,\alpha^2\,m_{h^0/H^0}^3}{128 \sqrt{2}\pi^3}\left\vert \sum_{f}\,N_c\,Q_f^2\,\lambda_{ff}^{h^0/H^0}\,A_{1/2}\left(\frac{m_{h^0/H^0}^2}{4\,m_f^2}\right)+\lambda_{WW}^{h^0/H^0}A_{1}\left(\frac{m_{h^0/H^0}^2}{4\,m_W^2}\right) \right.\nonumber \\
&-&\left.\frac{v^2}{2\,m_{H^+}^2}\lambda_{H^+\,H^-}^{h^0/H^0}\,A_0\left(\frac{m_{h^0/H^0}^2}{4\,m_{H^\pm}^2}\right)\right\vert^2 \,.
\eea
\ew
The first line is the contribution of SM fermions and $W$ boson running in the loop, while the second line contains the contribution from $H^\pm$. 
 As we said, the couplings of $h^0$ to SM fermions and gauge bosons are equally rescaled,  $\lambda_{ff}^{h^0}=\lambda_{VV}^{h^0}=\cos(\theta)$, while we have $\lambda_{H^+\,H^-}^{h^0}~\simeq~-(\kappa_{12}\,\cos(\theta)+\kappa_{23}\sin(\theta)\,v_\phi/v_1)$. 
Depending on the sign of the  coupling $\lambda_{H^+\,H^-}^{h^0}$, the diphoton rate $h^0 \to \gamma \gamma$ can be enhanced ($\lambda_{H^+\,H^-}^{h^0}>0$) or suppressed ($\lambda_{H^+\,H^-}^{h^0}<0$)~\footnote{Enhancement is also possible for large negative values of $\lambda_{H^+\,H^-}^{h^0}$, such that $H^\pm$ contribution largely dominates over SM ones. We do not consider this possibility.}, the effect being more pronounced for light $m_{H^{\pm}}$.

\subsubsection{Analysis}

We assume $m_{h^0}= 126$ GeV and $m_{h^0} < m_{H^0} \lesssim 600$ GeV.~\footnote{We only dispose of data up to $600$ GeV, that we use as an upper bound on the scalar masses.}
As previously said, the other neutral scalar fields $h_A$ and $A^0$ are very weakly coupled to the SM particles: their contributions can be neglected.
 
To perform the analysis, we construct the Higgs signal strengths $\mu_i$ corresponding to the five channels $h^0 (H^0)\to bb,\,\tau \tau, \gamma \gamma,\, WW$ and $ZZ$. 
 For the $b$ and $\tau$ channel, we use the combined results of CMS at 7 and 8 TeV~\cite{CMSB,CMSTau} and the 7 TeV results of ATLAS~\cite{ATLASB,ATLASTau}, while for the gauge boson channels we use both ATLAS and CMS results combining 7 and 8 TeV observations~\cite{ATLASGamma,ATLASZ,ATLASW,CMSGamma,CMSZ,CMSW}. 
We do not distinguish the decay products of the gauge bosons produced in $h^0/H^0$ decays, $h^0/H^{0} \to V V$, ($V=W^{(*)}, Z^{(*)}$): for both $h^0$ and $H^0$, the resulting branching factors cancel out with the SM ones. For $h^0$ we use combined results, while for $H^0$ only the most constraining signal at a given mass is considered.
 Further, for $h^0$  we combined ATLAS and CMS results assuming a Gaussian distribution for the observed signal strengths $\hat{\mu_i}$ and symmetric errors: we summarize in Table~\ref{data} the central value and the symmetric errors for the different channels we use. 
  \begin{table}[!t]
\begin{center}
\begin{tabular}{|c||c|c|c|c|c|}
\hline
\rule[0.15in]{0cm}{0cm}{\tt Channel:}& $\tau\,\tau$ & $b\,b$ & $W\,W$ &$Z\,Z$ & $\gamma\,\gamma$ \\
\hline
\rule[0.25in]{0cm}{0cm}$\hat{\mu}_i$ & $0.15$ & $0.49$& $0.9$ &$0.88$ & $1.67$ \\\hline
\rule[0.25in]{0cm}{0cm}$\sigma_i$ & $0.7$ & $0.73$& $0.3$ & $0.34$ & $0.34$ 
\\\hline
\end{tabular}
\caption{Best fit value $\hat{\mu}_i$ and symmetric errors given at 1 $\sigma_i$ used in our fit.}\label{data}
\end{center}
\end{table}%
More refined analyses have been done in e.g.~\cite{Giardino:2012ww,Espinosa:2012vu,Espinosa:2012im,Carmi:2012in,Carmi:2012zd}. We nevertheless do not expect that a more rigorous treatment of LHC data  would yield significant deviations from the results we obtain.
 
 Finally, we take into account the electroweak precision data. We construct the $S$, $T$ and $U$ parameters following the results of~\cite{Grimus:2008nb}, and used the values of the electroweak fit quoted in~\cite{Espinosa:2012im}:
 \bea
 S=0.0\pm 0.1\,,\quad T=0.02\pm 0.11\,,\quad U=0.03\pm 0.09\nonumber\,.
\eea

We then compute a $\chi^2$ defined by
\bea
\chi^2(\mu_i(h^0))=\sum_{i=\gamma,Z,W,S,T,U} \frac{(\mu_{i}(h^0)-\hat{\mu_i})^2}{\sigma_i^2}\,,\nonumber
\eea
that we minimize over the $\gamma$, $Z$ and $W^\pm$ Higgs signal strengths and the three oblique parameters.
As no significant excess of events has been seen so far in $b$ and $\tau$ channels, we do not include them in the definition of the $\chi^2$: we instead require that  $\mu_b$ and $\mu_\tau$ are below their respective 95\% C.L. upper bound. 
 We further ask the $H^0$ signal strength $\mu_i(H^0)\,(i=b,\tau, \gamma, Z, W)$  to be smaller than the observed  ones over the full $H^0$ mass range. 
 
\subsubsection{Results}

Before addressing the results, a few comments are in order.
The observables we consider are built upon a rich scalar sector: 9 free parameters are scanned over.
 We do not aim to constrain these parameters, but to highlight the main features our model exhibits.

 Constraints on the invisible decay width can already be set:
  in the SM, the branching ratio ${\rm Br}(h\to {\gamma\,\gamma})$ of a $126$ GeV Higgs is about $2.3\times10^{-3}$~\cite{HiggsWidth} and even if the diphoton rate is increased in our case, we can neglect it in a first approximation.
  Then, the total decay width of $h^0$ is approximately the sum of the rescaled SM channels plus the invisible one:
  \bea
  \Gamma(h^0)_{\rm tot}\simeq \cos(\theta)^2\,\Gamma(h)_{\rm tot}^{\rm SM}+\Gamma(h^0\to {\rm inv})\,.
  \eea 
  For the 4 tree level channels $h^0 \to b\,b$, $\tau\,\tau$, $W\,W$ and $Z\,Z$, we can write:
  \bw
  \bea
  \mu_i(h^0)=\frac{\sigma(pp\to h^0)_i\times {\rm Br}(h^0 \to i)}{\sigma(pp\to h)_i^{\rm SM}\times {\rm Br}(h \to i)^{\rm SM}}=\cos^4(\theta)\frac{\Gamma(h)_{\rm tot}^{\rm SM}}{\Gamma(h^0)_{\rm tot}}\simeq \cos^2(\theta)\,(1-{\rm Br}(h^0\to {\rm inv}))\,.
  \eea
  \ew

  Asking for example that the $h^0 \to W\,W$ signal strength is within its 2 $\sigma$ range, we have approximately an upper bound on the invisible width ${\rm Br}(h^0\to {\rm inv}) \lesssim 0.69$. The global fit to the 6 observables will however provide a slightly different bound.
  
\begin{figure}[t!]
\vspace{-2mm}{\includegraphics[width=0.48\textwidth]{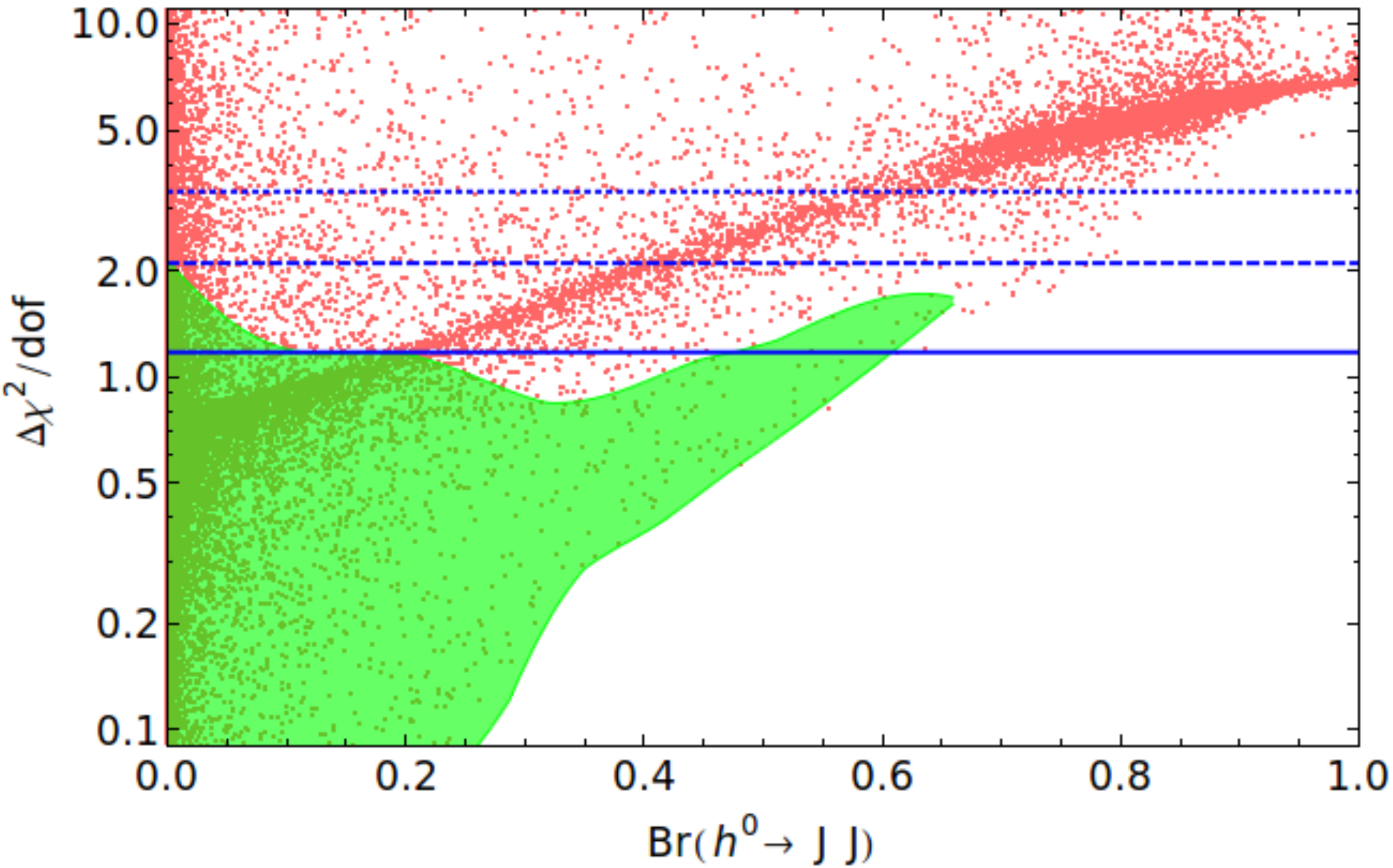}}\quad
\includegraphics[width=0.45\textwidth]{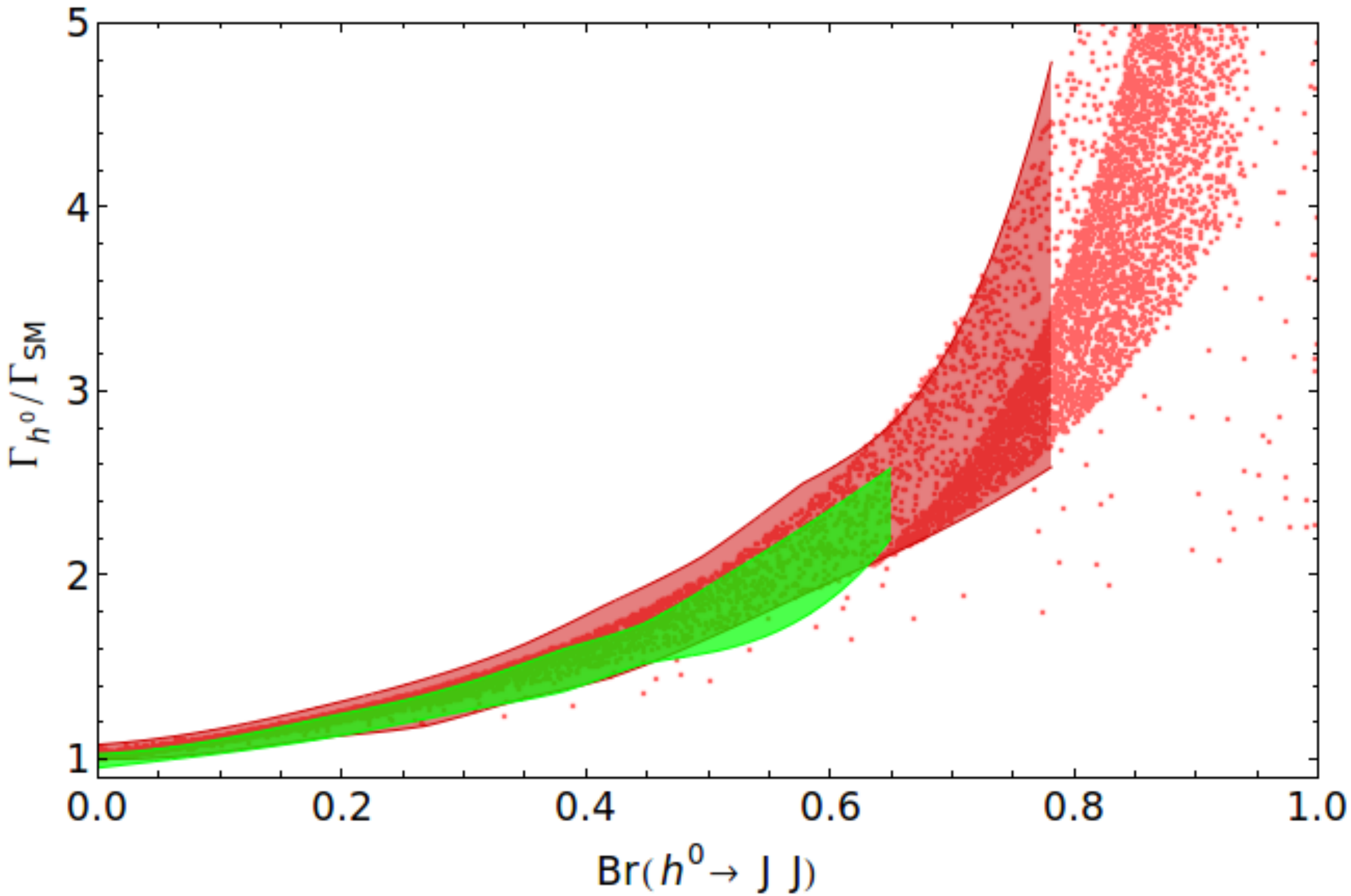}
\caption{\textbf{Left}: $\Delta\chi^2$ per degree of freedom as a function of $h^0$ invisible branching ratio. Blue solid-dashed-dotted lines represent 1-2-3 $\sigma$ deviations from the best fit value, respectively. The green area represents values for which all observables are within (below) their respective 95\% C.L. range (upper bound). \textbf{Right}: Total $h^0$ decay width normalized to the SM case as a function of the invisible branching ratio. The  shaded dark red region corresponds to the 95 \% C.L. range of the global fit, while the green area stands for observables  within (below) their respective 95\% C.L. bounds (upper bound).}
\label{fig:BrJJ}
\end{figure}

  We display in the left panel of Fig.~\ref{fig:BrJJ} the influence of the invisible branching ratio $h^0 \to J J$ on the goodness of the fit. The thin red dots are the values of $\Delta\chi^2=\chi^2-\chi^2_{min}$  per degree of freedom~\footnote{The minimal $\chi^2$ we obtain is $\sim 0.3$ for $6$ degrees of freedom.}, and the blue lines represent the $1$-$2$-$3$ $\sigma$ deviations from the best fit, from bottom to top. The green shaded area represents the allowed invisible branching ratio asking each $\mu_i$ to be within its 95\% C.L. allowed range  
  (but for $\mu_b$ and $\mu_\tau$ channels which are only upper bounded). At 95\% C.L., we obtain ${\rm Br}(h^0\to {\rm inv}) \lesssim 0.77$.
This value is large, and only possible from the increase  of the total $h^0$ decay width. In the right panel of Fig.~\ref{fig:BrJJ}, we show how  much this width is enlarged compared to the SM case, 
$\Gamma(h)_{\rm tot}^{\rm SM}\simeq\,4.2$ MeV~\cite{HiggsWidth}, when the invisible channel $h^0 \to J J$ increases.  Results of our scan are depicted by red dots, and we shade in dark-red the region within the 95\% C.L. allowed range of the global fit. Similarly, the green area represents the 95\% C.L. allowed range of each $\mu_i$ taken separately. Clearly, larger ${\rm Br}(h^0\to {\rm inv})$ are only possible for larger total $h^0$ width.
\begin{figure}[!t]
\includegraphics[width=0.6\textwidth]{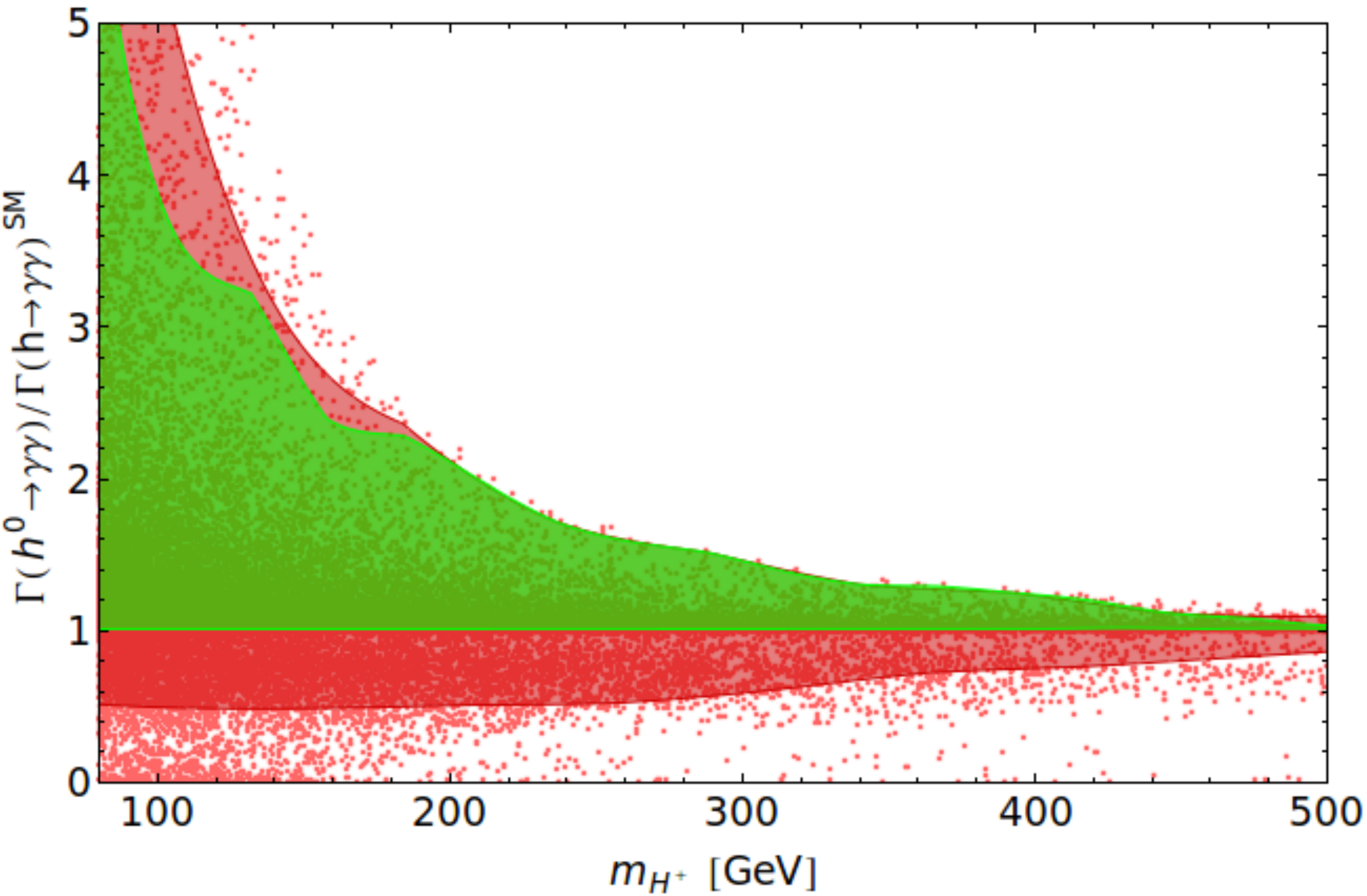}
\caption{Ratio $\Gamma(h^0 \to \gamma\, \gamma)$ to the SM rate in function of the charged Higgs mass $m_{H^\pm}$. Dark red region covers  the 95\% C.L. allowed range of the global fit, while for the green region  all the observables are within their respective 95\% C.L. bounds (or upper bounds for $b$ and $\tau$ final states).}
\label{fig:hgg}
\end{figure}
Another important result is the large possible enhancement of the diphoton rate. We display in Fig.~\ref{fig:hgg} the $h^0\to \gamma\,\gamma$ decay rate normalized to the SM one, showing the possible enhancement (or suppression) of this rate in function of the charged scalar $H^\pm$ mass. A large enhancement compared to the SM case is possible, although still compatible with current observations~\cite{ATLASGamma,CMSGamma}.

As a last comment, it is clear that our model can mimic the SM Higgs sector, or alternatively significantly deviate from it.
Further data will hopefully shed light on the nature of the observed $~126$ GeV excitation.

%%%%%%%%%%%%%%%%%%%%%%%%%%%%%%%%%%%%%%%%%%%%%%%%%%%%%
%%%%%%%%%%%%%%%%%%%%%%%%%%%%%%%%%%%%%%%%%%%%%%%%%%%%%
%%%%%%%%%%%%%%%%%%%%%%%%%%%%%%%%%%%%%%%%%%%%%%%%%%%%%
\section{Scalar Triplet Dark Matter}\label{Section:DM}

In our model the dark matter is the complex scalar field $S$ triplet of $\rm SU(2)_{\rm L}$, for which the most general scalar potential reads
\bw
\bea \label{VDM}
\mathcal{V}_{\rm DM}&=& \mu_{S}^2\, S^{\dagger} S + \lam_{S}\, (S^{\dagger} S)^2 + 
  \lam_{S}^\prime \left( S^{\dagger}  T_{G}^{a} S \right) \left( S^{\dagger }T_{G}^{a} S \right)+\Fs_{1}\, H_1^{\dag}\,H_1 S^{\dagger} S +\Fs_{2}\, H_2^{\dag}\,H_2 S^{\dagger} S + \Fs_{3}\, \phi^{*} \phi S^{\dagger} S\nonumber \\
&+&  \Fs_{1}^{\prime} \left( H_{1}^{\dagger} T_{2}^{a} H_{1}\right) \left( S^{\dagger } T_{G}^{a} S \right)+ \Fs_{2}^{\prime} \left( H_{2}^{\dagger} T_{2}^{a} H_{2}\right)\left(S^{\dagger} T_{G}^{a} S \right)+\mathcal{H}\, S^2 H_1^{\dag}\,H_2 + \mathcal{H}^{*}\, S^{\dagger 2} H_2^{\dag}\,H_1 \nonumber \\
 &-& \frac{\mupp}{\sqrt{2}} (S^2 \phi^{*} + S^{\dagger 2} \phi)\,.
    \eea
    \ew
  We assume without loss of generality a real $\mathcal{H}$ coupling, while the phase on $\mupp$ has been rotated away.
 In the definition  above  $T_{2}^{a}$ ($T_{G}^{a}$)  denote the three generators of $\rm SU(2)_{\rm L}$ in the fundamental (adjoint) representation.
 Notice that for a real triplet, the  terms $\left( S^{\dagger}  T_{G}^{a} S \right)$ identically vanish. The components of the triplet in the Cartesian basis are $S=(S^{1},~S^{2},~S^{3})$, the $S^{a}$ ($a=1,2,3$) being complex. In the spherical basis, the $S$ field reads:~\footnote{The unitary transformation from the Cartesian to the spherical basis is: $S\to U^{\dagger} S$, with $S=(S^{1},~S^{2},~S^{3})$ and 
 \begin{equation}
 U=\frac{1}{\sqrt{2}}\left(
\begin{array}{ccc}
1 & 0 & 1\\
i & 0 & -i\\
0 & \sqrt{2} & 0
\end{array}
\right)\,.	\nonumber
 \end{equation}
 In the spherical basis $S=(S^{+},~S^{0},~S^{-})$ the three generators of  $\rm SU(2)_{\rm L}$ in the adjoint representation are:
 \begin{eqnarray}
 	T_{G}^{1}=\frac{1}{\sqrt{2}}\left(
\begin{array}{ccc}
0 & -1 & 0\\
-1 & 0 & 1\\
0 & 1 & 0
\end{array}
\right)\,,\quad
T_{G}^{2}=\frac{1}{\sqrt{2}}\left(
\begin{array}{ccc}
0 & i & 0\\
-i & 0 & -i\\
0 & i & 0
\end{array}
\right)\,,\quad
T_{G}^{3}=\left(
\begin{array}{ccc}
1 & 0 & 0\\
0 & 0 & 0\\
0 & 0 & -1
\end{array}
\right)\,.\nonumber
 \end{eqnarray} 
Notice that in the spherical basis the potential given in (\ref{VDM}) is still invariant under the symmetries of the Lagrangian, provided
one replaces the operators $S^{2}$  with $S\, T_{3}\, S $, where the transformation matrix $T_{3}$ is given by:
\begin{eqnarray}
T_{3}=\left(
\begin{array}{ccc}
0 & 0 & 1\\
0 & 1 & 0\\
1 & 0 & 0
\end{array}
\right)\,.\nonumber
\end{eqnarray} } 
  $S=(S^{+},~S^{0},~S^{-})$, with $S^{\pm}\equiv(S^{1}\mp i S^{2})/\sqrt{2}$
 and $S^{0}\equiv S^{3}$.  Notice that, since the Cartesian components of $S$ are complex fields, $(S^{\pm})^* \neq S^{\mp}$.
  The triplet S thus describes 6 real degrees of freedom and after symmetry breaking the  spectrum consists of 4 massive particles: two neutral real scalars $S_{L/H}^0$ and two charged  $S_{L/H}^\pm$, the subscript denoting the lightest/heaviest mass eigenstate.
 Hence, the components of $S$ can be written as:
 \bw
\bea
S=\left( \cos(\theta_s)\,S_{L}^{+}+\sin(\theta_s)\,S_{H}^{+},\,\left(S_L^0 + {\rm i } S_H^0\right)/\sqrt{2}\,,\, \cos(\theta_s)\,S_{H}^{-}\,-\sin(\theta_s)\,S_{L}^{-} \right)\,.
\eea
\ew
Defining
\bea
m_0^2&=&\mu_S^2+\left( \Fs_1\,v_1^2+\Fs_2\,v_2^2+\Fs_3\,v_\phi^2\right)/2\,,\nonumber \\ 
\delta_0^2 &=& 2\mupp\,v_\phi-2\,\mathcal{H}\,v_1\,v_2\quad {\rm and }\quad\delta_\pm^2=(\Fs_1^\prime\,v_1^2+\Fs_2^\prime\,v_2^2)/2, \nonumber
\eea
the mixing angle is $2\,\theta_s=- {\rm Arctan}( \delta_0^2/\delta_\pm^2)$ and the mass spectrum is at tree level: 
\bea\label{DMmass}
m_{S_{L(H)}}^{0}&=&\left(m_0^2\mp \frac{\delta_0^2}{2}\right)^{1/2}\,,\nonumber \\
 m_{S_{L(H)}}^{\pm}&=&\left(m_0^2\mp \frac{1}{2}\sqrt{\delta_0^4+\delta_\pm^4}\right)^{1/2}\,.\label{massDM}
\eea
These expressions are valid as long as $\sqrt{\delta_0^4+\delta_\pm^4} \leq 2\,m_0^2$.

\begin{figure}[t!]
\begin{center}
\includegraphics[width=0.6\textwidth]{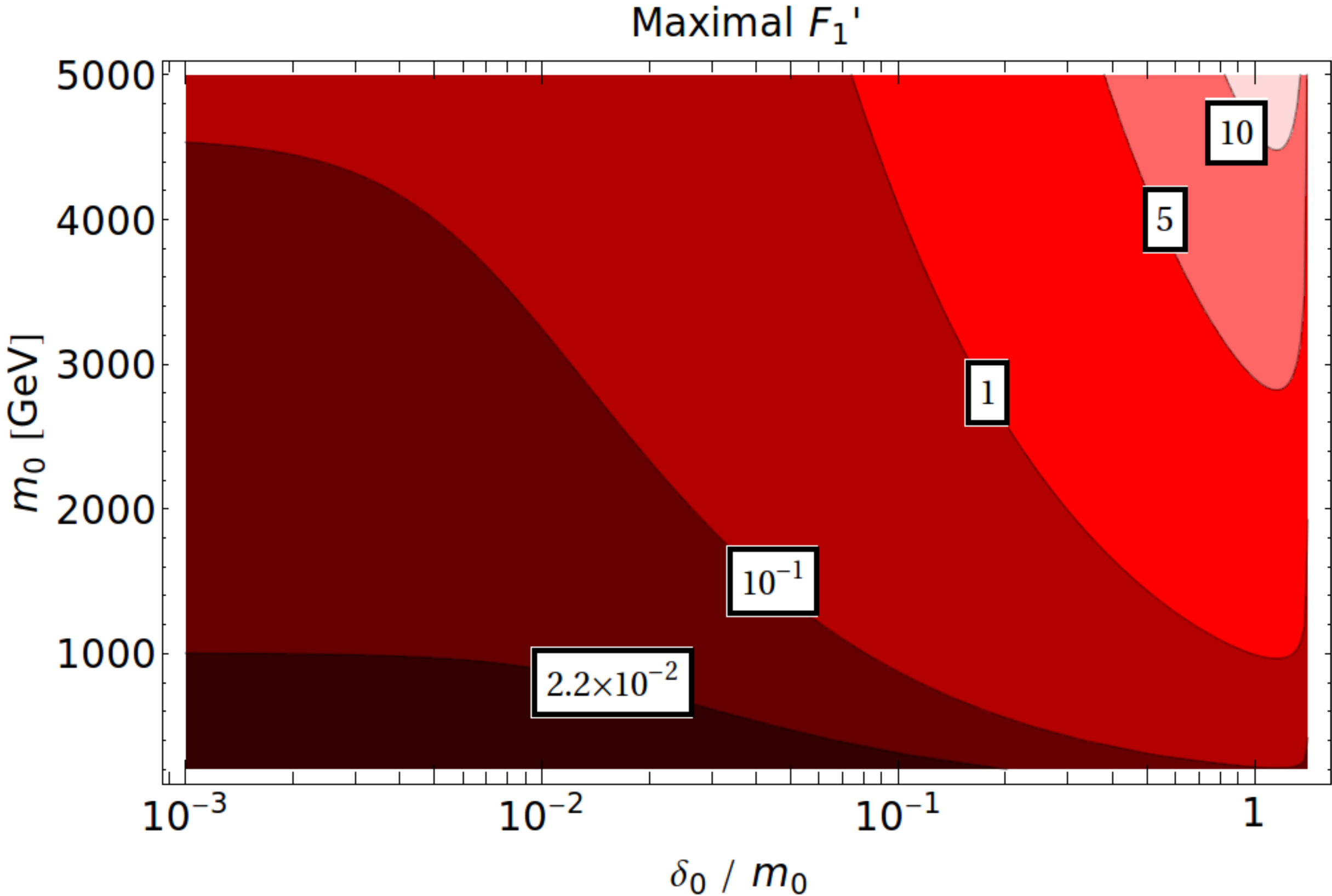}
\caption{Maximal value of the coupling $\Fs_1^\prime$ in the $\delta_0/m_0$ vs $m_0$ plane, such that $m_{S_L}^0< m_{S_L}^\pm$ .}
\label{fig:MaxFp1}
\end{center}
\end{figure}
At tree level, the  charged fields receive a mass contribution from $\delta_\pm$. As a consequence, the lightest charged scalar, $S_L^\pm$, 
is lighter at tree level than the lightest neutral component, $S_{L}^0$. However, at one loop, the charged-neutral mass splitting receives a positive contribution from gauge boson loops~\cite{MinimalDM}:

\bea
\left(m_{S_L(H)}^{\pm}-m_{S_L(H)}^{0}\right)_{\vert 1 {\rm loop}}=\left(m_{S_L(H)}^{\pm}-m_{S_L(H)}^{0}\right)_{\vert{\rm tree}}+\delta_m\,,\nonumber
\eea

with $\delta_m \simeq (166\pm 1) \MeV$ in the triplet case.
Thus, $S_L^0$ constitutes a viable DM candidate if $\delta_\pm$ contribution is smaller than $\delta_m$. In the case $\delta_0=0$ we have:
\bea
\Fs_1^\prime \lesssim 2.2\times 10^{-2}\,\left(\frac{m_0}{1 \TeV}\right)\,,
\eea
a similar bound derived in~\cite{Hambye:2009pw}.~\footnote{The case discussed in~\cite{Hambye:2009pw} actually corresponds to $\delta_0^2=0$ and $v_2=v_\phi=0$.}
However, if $\delta_0\neq 0$, this upper limit can be relaxed, as we show in Fig.~\ref{fig:MaxFp1}. When $\delta_0$ is comparable to $m_0$ (still smaller than $\sqrt{2}\,m_0$), $\Fs_1^\prime$ can take sizable values.  Under these conditions, $S_L^0$ provides a viable DM candidate. 

\subsection{Relic abundance}

The study of the scalar triplet as a possible DM candidate has been investigated in detail in e.g.~\cite{MinimalDM,Hambye:2009pw}.
 Our results agree with theirs.
  The dominant annihilation channel depends on the DM mass: the lowest mass is reached  in the pure gauge boson limit, when Higgs portals are negligible.  $S$ annihilations into gauge bosons receive in this case contributions either from the quartic coupling $\propto g\,S\,S^\dagger\,A_\mu\,A^\mu$ or through the trilinear coupling $\propto g\,S^\dagger S\,A^\mu$. For heavier DM, additional annihilation channels are mandatory in order to get sufficiently large annihilation cross sections. In our case $S$ can annihilate into the scalars $H_{1,2}$ and $\phi$ through the Higgs portal couplings $\Fs_i$ and $\Fs_i^\prime$.
Owing to the scalar potential  we considered, our model encompasses the cases discussed in~\cite{MinimalDM,Hambye:2009pw}.
The case of a complex triplet can be recovered if we enforce $\delta_0=0$, while the case $\delta_\pm=0$ resembles the real case (as the term $\Fs_1^\prime$ would identically vanish in such case), although we have twice more degrees of freedom.

When the mass parameter $\delta_\pm$ in eq.~(\ref{massDM}) is suppressed compared to both $\delta_0$ and $m_0$, the mass splitting between the neutral and the charged component of each pair $m_{S_L}$/$m_{S_H}$ is negligible.
 Coannihilations between neutral and charged components become, then, important. 
 In the regime $\delta_\pm \ll \delta_0,\,m_0$, if $\delta_0$ is comparable to $m_0$, $m_{S_L}^0 \sim m_{S_L}^\pm$ is much lighter than $m_{S_H}^0 \sim m_{S_H}^\pm$ and a lower bound $m_{S_L}^0 \gtrsim 1.8$ TeV is then obtained~\cite{MinimalDM,Hambye:2009pw}. If, oppositely, $\delta_0 \ll m_0$, the number of annihilating particles effectively double and the lower bound $m_{S_L}^0 \gtrsim 1.8/\sqrt{2}$ TeV is reached.

\begin{figure}[t!]
\begin{center}
\includegraphics[width=0.48\textwidth]{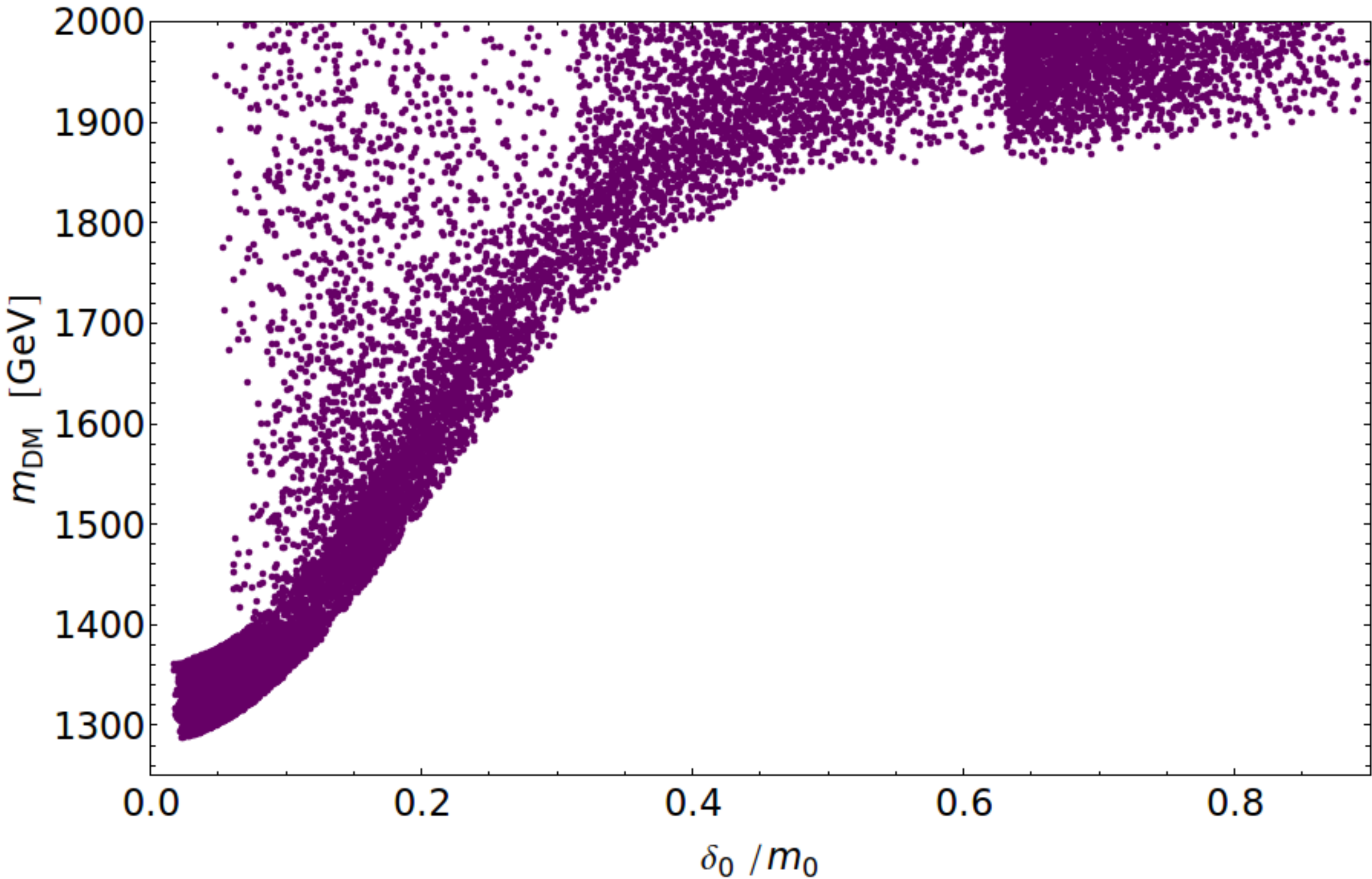}\,\quad\includegraphics[width=0.48\textwidth]{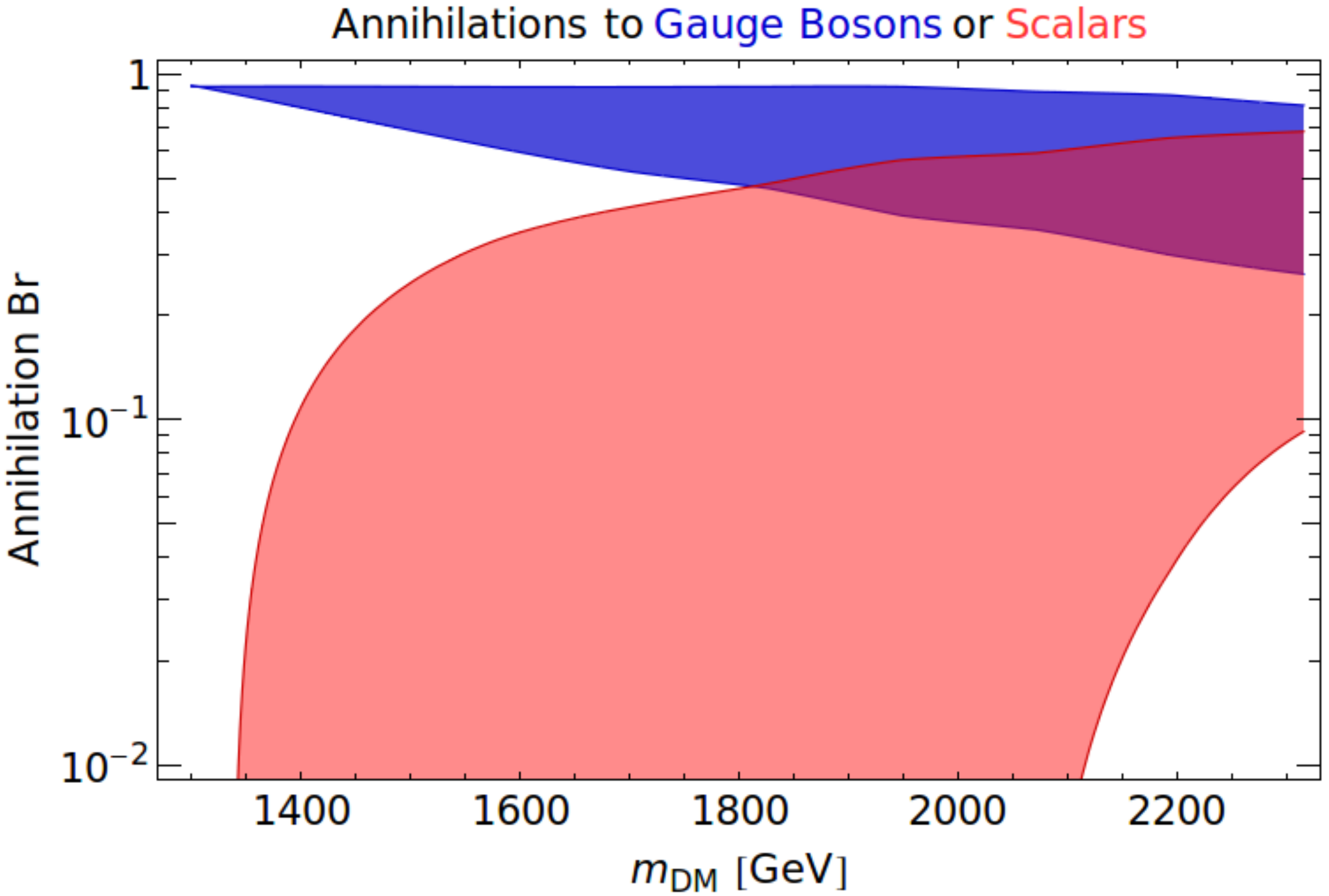}
\caption{{\bf Left}: dark matter mass, $m_{\rm DM}$, as function of the mass parameter $\delta_0/m_0$ (see the text). {\bf Right}: dark matter annihilation branching ratio versus  $m_{\rm DM}$: in blue are shown the annihilations into gauge bosons while the red area corresponds to  annihilations into scalars.}
\label{fig:DM1}
\end{center}
\end{figure}

To verify the validity of $S_{L}^0$ as a DM candidate, we implement our \textit{complete} model in FeynRules~\cite{FeynRules} to generate the CalcHep~\cite{Pukhov:2004ca} files used for micrOMEGAs~\cite{micromegas}. We then scan over 19 parameters~\footnote{9 free parameters for the symmetry breaking potential and 10 for the DM sector. The fermion sector is fixed.}, and demand the DM relic density to lie within the 1$\sigma$ range~\cite{PDG12}: $\Omega_{\rm DM}h^2=0.112\pm 0.006$. We found, numerically, the following lower bound at 1 $\sigma$:
\bea
m_{\rm DM} \gtrsim 1290\,\GeV\,,
\eea
in agreement with~\cite{Hambye:2009pw}.
 As remarked above, this lower bound is obtained when the 4 $Z_{2}-$odd scalars contribute to the relic density, i.e. in the regime of low splitting $\delta_0 \ll m_0$.
  We illustrate this in the left panel of Fig.~\ref{fig:DM1}, where the values of $m_{\rm DM}$ are plotted against the ratio $\delta_0/ m_0$, recalling eq.~(\ref{DMmass}).
   When $\delta_\pm\lesssim \delta_0 \ll m_0$, the DM mass reaches its minimum, while when $\delta_0$ becomes comparable to $m_0$, the lowest DM mass is about $m_{\rm DM}\sim 1860$ GeV, the lowest values reached for small $\delta_\pm$ (corresponding to the case $\lambda_3=0$ of~\cite{Hambye:2009pw}).
    We also show, in the right panel of Fig.~\ref{fig:DM1}, the range the annihilating branching ratios span as function of $m_{\rm DM}$.
     In blue are depicted the annihilations $S\,\bar{S}\to V\,V$ ($V=W,Z$), while in red we show the branching fraction for $S\,\bar{S}\to \phi_i\,\phi_j$, with $\phi_i$ any of the $Z_2-$even scalars present in our model.
      Over the mass range depicted, annihilations to gauge bosons mostly fix the DM relic abundance, as explained in~\cite{MinimalDM,Hambye:2009pw}. However for $m_{\rm DM}\gtrsim 2$ TeV, the scalar contribution is necessary to compensate the $m_{\rm DM}^{-2}$ suppression of the annihilation cross section.

\subsection{Probing a scalar triplet dark matter}
Probing the triplet nature of the DM could  be achievable, in principle, in direct and indirect detection experiments, and in collider searches.
The indirect detection would consist in the observation of charged cosmic rays; however, the predicted fluxes are too suppressed compared to the background to be measured~\cite{MinimalDM,Hambye:2009pw}.
 Direct detection is more promising, and forthcoming experiment XENON 1T~\cite{XENON1T} can probe a part of the parameter space, corresponding to the larger values of $\Fs_1^{(\prime)}$.
As for colliders, the case for triplet scalars have been studied in~\cite{MinimalDM,ScalarLHC}.
 In our case with complex fields, the heavy particles decay to the light one plus a Majoron,  $S_{H}^{0(\pm)} \to S_{L}^{0(\pm)}+J$, while the lightest charged scalar $S_L^\pm$ decays almost exclusively to $S_L^\pm \to S_L^0+\pi^{\pm}$.
  However, these channels provide no hope for probing the triplet nature of DM~\cite{MinimalDM,ScalarLHC}.
Therefore, only if the quartic couplings $\Fs_1^{(\prime)}$ are large enough one can hope to observe a characteristic signal of the triplet scalar $S$.

\section{Neutrino Masses and Leptogenesis}\label{Sect3}

Along the lines of~\cite{Previous}, we explain the light neutrino masses within an inverse see-saw realization~\cite{Mohapatra:1986bd}, using a minimal field content.
In this scenario, a vectorlike fermion, singlet under the SM gauge group is introduced.
 This field has a  charge -$1$ under ${\rm U(1)_{\rm X}}$, a symmetry constructed to generalize the $B$-$L$ quantum number.
  By the addition of two scalar fields, the doublet $H_2$ and the singlet $\phi$ introduced in the previous sections, a consistent UV completion of the inverse see-saw is possible.
  
In the present scenario we introduce  a vectorlike Dirac fermion, $N_{D}$,  that is a triplet of SU$(2)_{\rm L}$ and is constructed from the RH neutrinos $N_{1,2}$,  
whose quantum numbers are reported in Table~\ref{FieldAssign}: $N_D \equiv P_R\,N_1+P_L\,N_2^C$.
  The interaction field $N_{D}$  couples to SM leptons via $H_{1,2}$ and $\phi$, and to the DM triplet scalar $S$. The couplings to leptons allow for an inverse see-saw generation of neutrino masses similar to the singlet case, and the coupling between $N_{D}$ and $S$ is necessary for the production of a baryon asymmetry in this model.
 
Indeed, as outlined in \cite{Previous} and further discussed below, an nonzero $N_{D}$-$\overline{N}_{D}$ number density is produced in the early universe if one postulates the existence of a SM singlet Majorana fermion, $N_{3}$, which  decays out of equilibrium in $N_{D}$ and $S$. This asymmetry is then transferred to SM lepton doublets through fast neutrino Yukawa interactions and partially converted into nonzero baryon number by in-equilibrium sphaleron nonperturbative processes.
\newline

Below we discuss the resulting neutrino mass spectrum and the main features of the leptogenesis mechanism.

\subsection{Type III inverse see-saw realization}
The most general interaction Lagrangian involving the Dirac field $N_{D}=(N_{D}^{1},\,N_{D}^{2},\,N_{D}^{3})$ (in the Cartesian basis) and  the Majorana singlet $N_{3}$ is
\bw
\bea
\mathcal{L}&\supset& -\,m_{N}\,\ovl{N_{D}^a}\,N_{D}^{a} - \left( Y_{\nu 1}^{\beta}\, \ovl{N_{D}^{a}} \,\widetilde{H}_1^{j\,*}\,(T_{2}^a)_{jk}\,\ell_\beta^{k}\,+ Y_{\nu 2}^{\gamma} \,\ovl{N_{D}^{a}}^{C}\, \widetilde{H}_2^{j\,*}
\,(T_{2}^a)_{jk}\,\ell_\gamma^{k}+\frac{\delta_N}{\sqrt{2}}\,\phi\,\ovl{N_{D}^a}\,N_{D}^{a\,C}  \,+\,{\rm h.c.}\right)\nonumber \\\label{neutrL}
&&-\frac{1}{2}\,M_{3}\,\overline{N_{3}}\,N_{3}^{C}\,-\, \left(h\, S^{a}\, \ovl{N_{D}^{a}}\, N_{3}  \,+\,{\rm h.c.}  \right)\,,
\eea
\ew
where $\ell_{\beta}=(\nu_{\beta\,\bf L},e_{\beta\,\bf L})^{T}$ ($\beta=e,\mu,\tau$), $N_{D}^{a\,C}\equiv C\overline{N}_{D}^{a\,T}$ ($a=1,2,3$) and $\widetilde{H}_{k}\equiv -i\sigma_{2}H_{k}^{*}$ ($k=1,2$). The parameter $\delta_N$ is made real through a phase transformation.
The first line of~\eqref{neutrL} contains terms providing masses to neutrinos after EWSB, while the second line contains additional terms required for leptogenesis. 
In the spherical basis of the adjoint representation, the components of $N_D$ are
\bw
\bea
N_D=\left(N_D^+\,,\;N_D^0\,,\;N_D^-\right)=\left(P_R\,N_1^++P_L\,(N_2^-)^C\,,\;P_R\,N_1^0+P_L\,(N_2^0)^C\,,\;P_R\,N_1^-+P_L\,(N_2^+)^C\right)\,,
\eea
\ew
and in particular $(N_D^{\pm})^C \neq N_D^{\mp}$, similarly to the complex scalar triplet $S$.

Notice that the Yukawa interactions $\propto Y_{\nu 1}$ ($Y_{\nu 2}$) couple $N_1$ ($N_{2}$) to the SM leptons. Therefore, 
after the Higgs doublets acquire a nonzero vev, the SM lepton number (i.e. the generalized $X$ charge) is explicitly violated by $Y_{\nu 2}$ mediated interactions.
Furthermore, while the Dirac type mass $m_{N}$ conserves the lepton number, 
the term $\propto \delta_N$ provides, after $\phi$ takes a nonzero vev, a Majorana mass term for both $N_1$ and $N_2$.
In the case $m_N\gg \delta_N\,v_\phi$ we have, in general, the inverse see-saw mechanism \cite{Mohapatra:1986bd}. 
Finally, as in usual type III see-saw scenarios~\cite{SeesawIII}, the charged components of the triplet and the SM leptons mix after EWSB, implying typically larger contribution to lepton flavour violation processes than the singlet RH neutrino case (see e.g.~\cite{Abada:2007ux,Abada:2008ea,Ibarra:2011xn,Dinh:2012bp}).\\

In the chiral basis $\left(\mathbf{\nu_{ L}}\;, (N_{1}^{0\,C})_{\bf L},\, (N_{2}^{0\,C})_{\bf L} \right)$, the $5\times 5$ symmetric neutrino mass matrix  reads:
\bea
\mathcal{M}_{\nu}=
\left(
\begin{array}{ccc}
\mathbf{0_{3\times3}} & \mathbf{y_{1}}\,v_1 & \mathbf{y_{2}}\,v_2\\
\mathbf{y_{1}}^{\rm T}\,v_1 & \delta_N\,v_\phi & m_{N}\\
\mathbf{y_{2}}^{\rm T}\,v_2 & m_{N} & \delta_N\,v_\phi
\end{array}
\right)\,.\label{massnu}
\eea
In eq.~(\ref{massnu}) $\mathbf{0_{3\times3}}$ is the null matrix of dimension 3 and we introduce the shorthand notation: $\mathbf{y_{k}}\equiv \left( Y_{\nu k}^{e}\;,Y_{\nu k}^{\mu}\;,Y_{\nu k}^{\tau}\right)^{\rm T}/2\sqrt{2}$ with  $k=1,2$.
The neutral spectrum, therefore, consists of one massless neutrino, two massive light Majorana neutrinos and two heavy Majorana neutrinos $\mathcal{N}_{1,2}$. These two particles are quasidegenerate, with mass $M_{\mathcal{N}_{1(2)}}=m_N\mp\ \delta_N\,v_{\phi}$, and form a pseudo-Dirac pair in the limit $m_N\gg \delta_N\,v_\phi$  \cite{PseudoDirac1,PseudoDirac2,PseudoDirac3}. 

The effective light neutrino mass matrix has elements
\bea
(M_\nu)^{\al \beta}\simeq -\frac{v_1\,v_2}{m_N}\,\left(\mathbf{y_{1}}^{\alpha }\,\mathbf{y_{2}}^{\beta}+\mathbf{y_{2}}^{\alpha }\,\mathbf{y_{1}}^{\beta}\right)+v_\phi\,\delta_N\,\frac{v_1^2}{m_N^2}\left(\mathbf{y_{1}}^\al \mathbf{y_{1}}^\be+\mathbf{y_{2}}^\al\, \mathbf{y_{2}}^\be \,\frac{v_2^2}{v_1^2}\right)\,.\label{massnu4}
\eea
This expression clearly highlights the different contribution to the light neutrino masses.
The first term acts as a linear seesaw, and its suppression originates from the small vev $v_2$.
The second term is typical of inverse seesaw scenarios,  
where the small ratio $v_\phi\,\delta_N / m_N$ suppresses the neutrino mass scale. 
We remark that, since only  two RH neutrinos $N_{1,2}$ are introduced, the linear seesaw contribution alone (i.e. neglecting $v_{\phi}$ in (\ref{massnu4}))
allows us to fit  all current neutrino oscillation data, while
if $v_{2}=0$  the inverse seesaw scenario can only account for one massive light neutrino.
Therefore, the complex scalar field $\phi$, with vev $v_{\phi}\neq 0$, in principle is not necessary in order to obtain two massive light neutrinos through the (linear) see-saw mechanism.
On the other hand, $v_{\phi}\neq 0$ is a necessary condition to set a hierarchy between the Higgs doublet vevs, $v_{2}\ll v_{1}$, without fine-tuning of the parameters \cite{Previous}.
Finally, we remark that the scalar field $\phi$ with a coupling $\delta_{N}\neq 0$ allows us to implement successful leptogenesis through the ``two-step'' scenario discussed below.

Using as shorthand notation $A=M_\nu\,M_\nu^\dagger$, the two nonzero neutrino masses are given by
\bea
m_\nu^{\pm}= \frac{1}{\sqrt{2}}\sqrt{{\rm Tr}(A)\pm\sqrt{2\,{\rm Tr}(A^{2})-{\rm Tr}(A)^2}}\,.\label{nmass1}
\eea
In terms of the see-saw parameters in the Lagrangian, we have
\bea
\label{TrA}
{\rm Tr}(A)&=&\frac{1}{m_N^2\,(1-\mu_N^2)^2}\times\left[2\,\left(y_1^2\,y_2^2+\vert y_{12} \vert^2\right)-4\,\mu_N\,(y_1^2+y_2^2)\,{\rm Re }(y_{12})+\mu_N^2\left(y_1^4+y_2^4+2\,\,{\rm Re }(y_{12}^2)\right) \right]\,,\\
\label{TrAA}2\,{\rm Tr}(A^{2})&-&{\rm Tr}(A)^2 = \frac{1}{\left(m_N^2(1-\mu_N^2)^2\right)^2}\times\,\left\lbrace \left[ 4\,\vert y_{12} \vert^2-4\,\mu_N\,(y_1^2+y_2^2)\,{\rm Re }(y_{12})+\,\mu_N^2\left(y_1^4+y_2^4+2\,\,{\rm Re }(y_{12}^2)\right)\right]\,\times \right. \nonumber \\
& &\left.\left[ 4\,y_1^2\,y_2^2 -4\,\mu_N\,(y_1^2+y_2^2)\,{\rm Re }(y_{12})+\,\mu_N^2\left(y_1^4+y_2^4+2\,\,{\rm Re }(y_{12}^2)\right)\right]+4\,\mu_N^2(2-\mu_N^2)\,\vert \eta_{12}\vert^2\,\right\rbrace,
\eea 
where we introduce for convenience  $y_i=\sqrt{\mathbf{y_{i}}^\dagger\cdot\mathbf{y_{i}}}\, v_i $, $y_{12}=\mathbf{y_{1}}^\dagger\cdot\mathbf{y_{2}} \, v_1\,v_2$, $\eta_{12}=\mathbf{y_{1}}\times \mathbf{y_{2}}\, v_1\,v_2$ and $\mu_N= (\delta_N\,v_\phi)/m_N$.
In the inverse seesaw limit, $\mu_N \ll 1$, the neutrino masses have a simple expression 
\bea
m_\nu^\pm &\simeq & \frac{1}{m_N}\left(\sqrt{ y_1^2\,y_2^2-\mu_N\,(y_1^2+y_2^2)\,{\rm Re }(y_{12})}	\pm \sqrt{ |y_{12}|^2-\mu_N\,(y_1^2+y_2^2)\,{\rm Re }(y_{12})}\right) \nonumber \\
&\simeq& \frac{1}{m_N}\left(\, y_1\,y_2\pm \vert y_{12}\,\vert\right)\times\left(1 \mp \frac{\mu_N}{2}\frac{(y_1^2+y_2^2)\,{\rm Re}(y_{12})}{y_1\,y_2\,\vert y_{12} \vert}\,\right).
\eea
Notice that if the neutrino Yukawa vectors  $\mathbf{y_{1}}$ and
 $\mathbf{y_{2}}$ are proportional, $m_{\nu}^{-}$ is exactly zero, in contrast with neutrino oscillation data. 
For a normal hierarchical spectrum, $m_\nu^+= \sqrt{\dma}$ and $m_\nu^-=\sqrt{\dmsol}$, while in the case of  inverted hierarchy 
we have $m_\nu^+=\sqrt{\dma}$ and $m_\nu^-=\sqrt{\dma-\dmsol}$, $\dma$ and $\dmsol$ being the atmospheric and solar neutrino mass square differences, respectively.
It is easy to show 
 that at leading order in $\mu_N$, the neutrino mass parameters satisfy the following relation
\cite{Previous}: $\left| \mathbf{y_{1} \times y_{2}}  \right|\,v_{1}\,v_{2}/m_{N} \cong (\dmsol\,\dma)^{1/4}$.
Hence, barring accidental cancellations, the size of
the neutrino Yukawa couplings is
\begin{equation}
|\mathbf{y_{1}}| |\mathbf{y_{2}}|\approx 10^{-8}\,(m_{N}/1~\text{TeV})\,(10~\text{MeV}/v_{2})\,.\label{yuk12}
\end{equation}

The model also accommodates two charged Dirac fermions, $\Sigma_{1,2}^{\pm}$, which at tree level are degenerate with the heavy neutral fermions $\mathcal{N}_{1,2}$. Similarly to the triplet scalar case,
the triplet gauge couplings induce a mass splitting $\simeq 166$ MeV \cite{MinimalDM} between the charged and neutral components of the triplet.\\

Production cross section at LHC is dominated by the triplet gauge interactions and scales with the overall mass of the triplet, $m_{N}$.  A discovery of these charged ($\Sigma_{1,2}^{\pm}$) and neutral ($\mathcal{N}_{1,2}$) fermions may be possible at LHC if $m_{N}\lesssim 1$ TeV \cite{typeIIILHC1,delAguila:2008cj,typeIIILHC2,typeIIILHC3,typeIIILHC4}. Current searches constrain the triplet mass to be above $m_N \gtrsim (180$-$210)$ GeV range~\cite{CMS:2012ra}.

\subsection{Leptogenesis mechanism}

 \begin{figure}[t!]
\begin{center}
\includegraphics[width=0.5\textwidth]{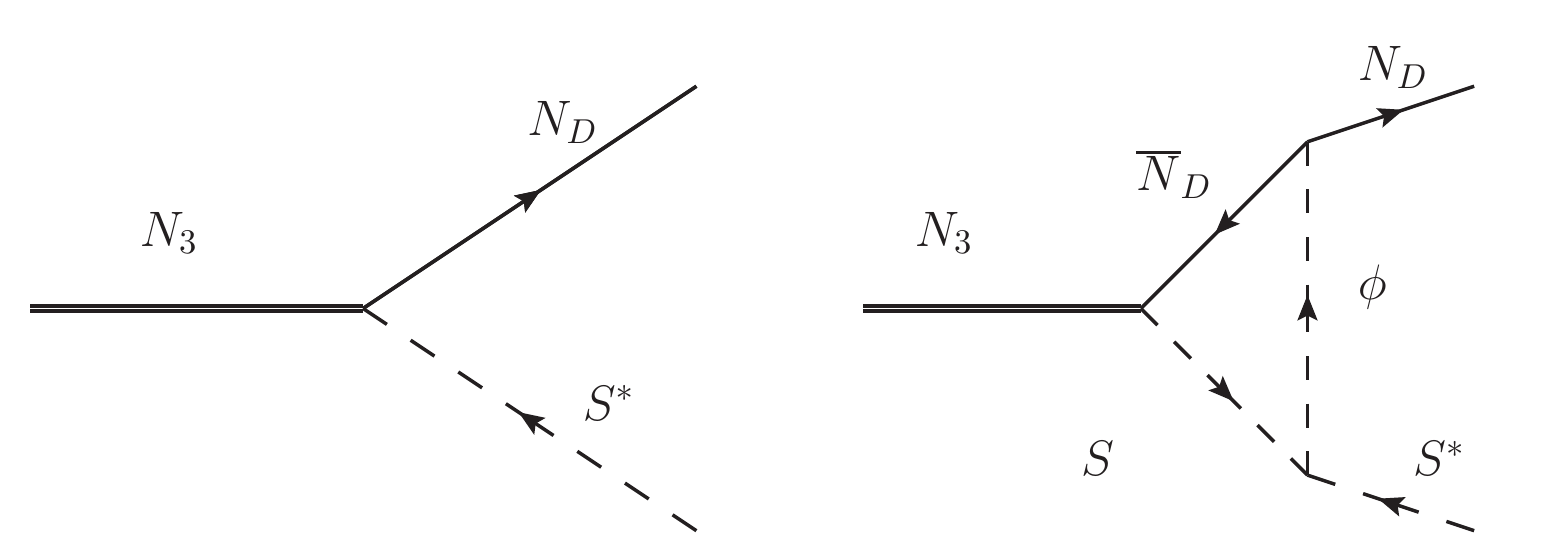}\
\caption{One-loop correction to $N_3$ decays allowing nonzero CP asymmetry.}
\label{Fig:eCPdiag}
\end{center}
\end{figure}

As anticipated at the beginning of this section, for a Majorana fermion $N_{3}$ heavier than the $\rm SU(2)_{\rm L}$ triplet fields $N_{D}$ and $S$, successful leptogenesis can be realized in this model within the ``two-step'' scenario introduced in \cite{Previous}:
\begin{itemize}

\item[\bf 1]  When the temperature of the universe decreases below  $M_{3}$,  $N_{3}$ out of equilibrium decays generate asymmetries in $S$ and $N_{D}$ abundances.
 A nonzero CP asymmetry, $\epsilon_{\rm CP}$, originates from the interference between the tree level and one-loop correction of $N_{3}$ decay amplitude \cite{Previous}
(see Fig.~\ref{Fig:eCPdiag}). We emphasize that the presence of $\phi$ with a coupling $\delta_N\neq 0$ in (\ref{neutrL}) is mandatory  in order to generate a nonzero $\epsilon_{\rm CP}$.
 The expression of $\epsilon_{CP}$ is in general lengthy, but shows a very simple dependence on the key parameters $\delta_N$, $h$ and $\mupp$ in the hierarchical limit $M_{3}^{2}\gg m_{N}^{2},\,m_{S}^{2}$:
\bw
\bea
\eps_{\rm CP}&\equiv& \frac{\sum_{a}\;\left[\Gamma\left(N_{3}\to N^{a}_{D}+\overline{S}^{a}\right)-\Gamma\left(N_{3}\to\overline{N}^{a}_{D}+S^{a}\right)\right]}
						{\sum_{a}\;\left[\Gamma\left(N_{3}\to\overline{N}_{D}^{a}+S^{a}\right)+\Gamma\left(N_{3}\to N_{D}^{a}+\overline{S}^{a}\right)\right]}
\nonumber\\
     &\simeq&- \frac{ \delta_N}{16\pi}\,\frac{{\rm Im}(h^2)}{\vert h \vert^2}\,\frac{\mupp}{M_3}\;\simeq\; -2\times 10^{-3}\,\delta_N\,\left(\frac{\mupp}{1\TeV}\right)\,\left(\frac{10\,\TeV}{M_3}\right)\,\frac{{\rm Im}(h^2)}{\vert h \vert^2}\,,
\label{ApprECP}
\eea 
\ew
\item[\bf 2] The neutrino Yukawa couplings mediate interactions between the SM leptons and $N_{D}$ which are in equilibrium at the leptogenesis temperature $T\sim M_{3}$. Such scattering processes, therefore, act as  a source term for a lepton number asymmetry, which is partially converted into a net baryon number by the fast sphaleron processes,  as in the standard thermal leptogenesis scenario. 
\end{itemize}

Besides asymmetry production and transfer processes, many washout interactions reduce the mechanism efficiency.
For the BAU yield to match observation, these processes should not too strongly deplete $N_{D}$ and lepton asymmetries.
In the singlet case~\cite{Previous}, we discussed in detail the corresponding region of the parameter space: typically, small values of $\vert h\vert \lesssim 10^{-4}$ reduce $N_3$ inverse decays and $\Delta N_D =1$ washout scatterings. These scatterings are similarly reduced for smaller values of $\delta_N$, although this in turn reduces $\epsilon_{\rm CP}$, hence larger $\mupp$ are needed in such cases.
In the present scenario, since both the fields $S$ and $N_{D}$ are taken in the adjoint of $\rm SU(2)_{\rm L}$,
 additional scatterings involving $S$, $N_{D}$, $N_{3}$ and the SM gauge bosons $V$ are typically very fast for a low see-saw scale $m_{N}$. The dominant washout processes, the $\Delta N_D=1$  ($N_{3}\, S \leftrightarrow N_{D}\, V$) and the $\Delta N_D=2$  ($N_{D}\,N_{D} \leftrightarrow \phi\, V$) (+ corresponding crossing symmetry), tend to strongly suppress the
 generated  lepton asymmetry.
All these interactions depend on the coupling $\delta_N$, which is then required to be small, smaller than in the singlet case.
Accordingly, $\mupp$ in the triplet case is typically larger than in the corresponding term in the singlet scenario \cite{Previous}, 
in order to obtain a sufficiently large $\epsilon_{\rm CP}$. Furthermore, the  scatterings with the gauge fields are controlled by the leptogenesis scale $T\sim M_{3}$, the see-saw scale $m_{N}$ and the dark matter dimensional parameter $\mu_{S}$.

We compute the final baryon asymmetry $Y_{B}$ within the present leptogenesis scenario for several values
of the relevant mass scales and couplings of the model and we compare it with the observed value~\cite{WMAP}: $Y_{B}^{\rm exp}=\left(8.77\pm 0.21\right)\times10^{-11}$.
 We solve for this a system of coupled Boltzmann equations for $N_{3}$, $N_{D}$, $S$ and
lepton abundance similar to the singlet scenario discussed in~\cite{Previous}, with the addition of the new interactions involving the gauge fields.

\begin{figure}[t!]
\begin{center}
\includegraphics[width=0.48\textwidth]{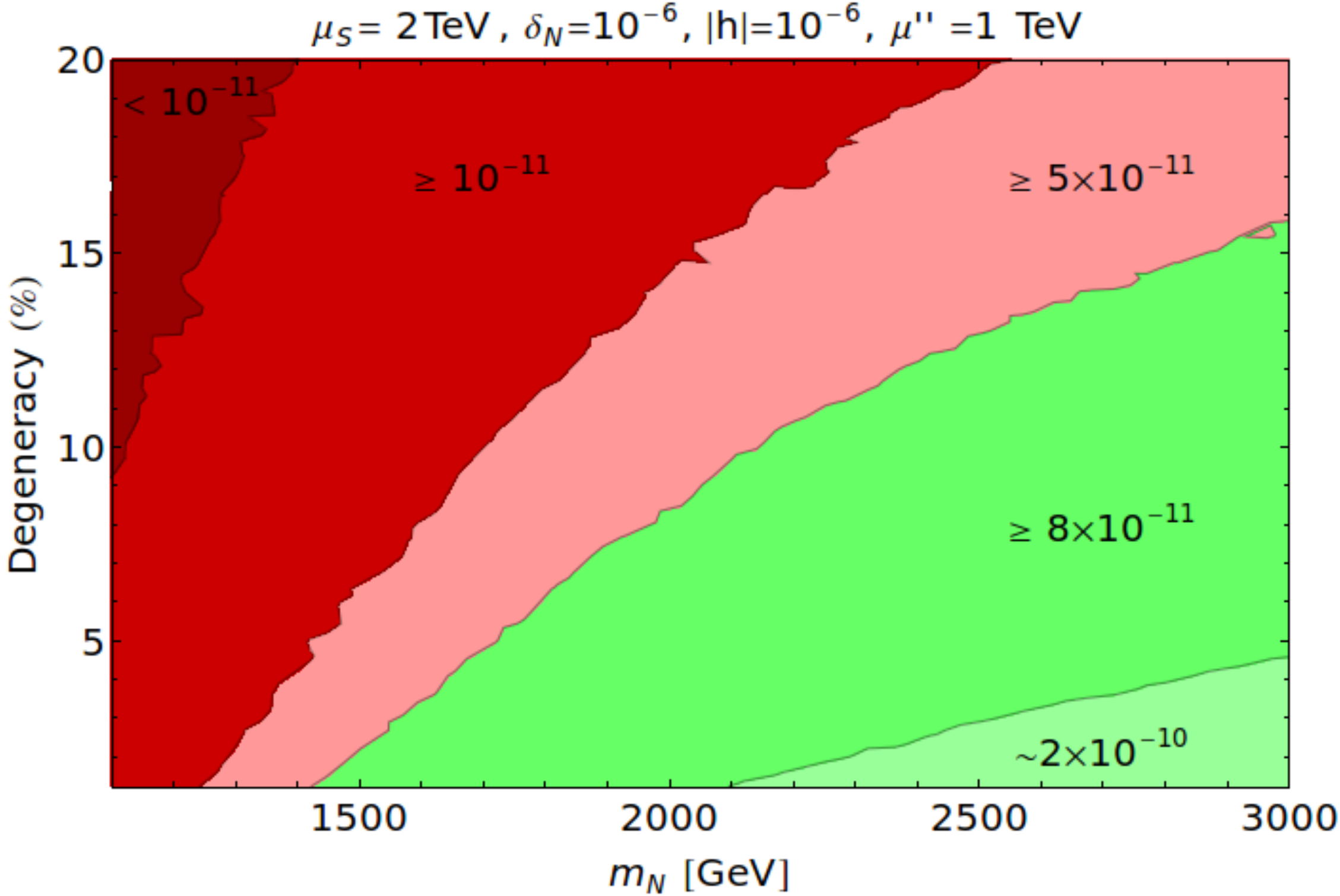}\; \includegraphics[width=0.48\textwidth]{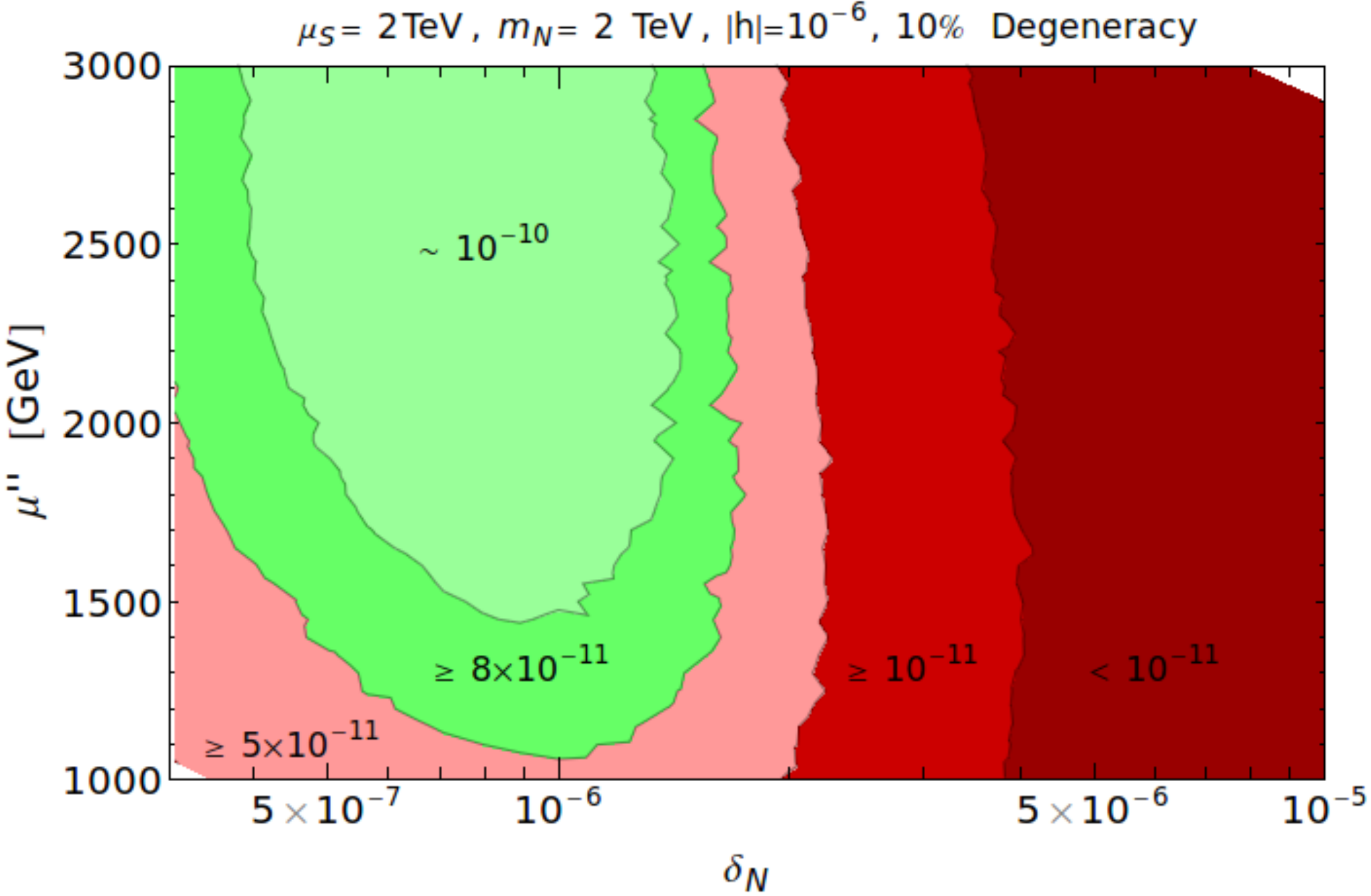}
\caption{{\bf Left:} Baryon asymmetry $Y_{B}$ as function of the degree of degeneracy between $M_3$ and $\mu_S+m_N$ and $m_N$ mass (see the main text). 
{\bf Right:} variation of $Y_{B}$ in terms of the parameters $\mupp$ and $\delta_{N}$, for fixed $M_3$, $\mu_S$ and $m_N$.  }
\label{figGammaRate}
\end{center}
\end{figure}

 Typical results of the numerical analysis are given in Fig.~\ref{figGammaRate}, where we show the variations of $Y_{B}$ against several key parameters. In both plots we fix for illustration $\vert h \vert=10^{-6}$ and $\mu_S=2$ TeV. The value of $h$ is about the maximal possible, while for $\mu_S=2$ TeV the DM mass is $m_{\rm DM}=(1930 \pm 70)$ GeV. 
 In the left panel of  Fig.~\ref{figGammaRate}, we highlight the dependence on  the see-saw scale $m_{N}$ and on $M_3$, through a degeneracy parameter equal to $1-(m_{N}+\mu_{S})/M_{3}$ (in percentage).
 In the singlet case we found no actual lower bound on $m_N$ from BAU requirement: clearly in the triplet case the picture changes.
 As is well known, in type III  leptogenesis, a lower bound on the see-saw scale $m_{N} \gtrsim 1.6$ TeV is set~\cite{typeIIIlepto2,typeIIIlepto3}.
 A similar bound is found here, although for very different reasons.
 In the standard type III case, gauge scatterings maintain fermion triplets in thermal equilibrium down to low temperatures, possibly after the decoupling of sphaleron processes, hence heavy enough triplets are required.
 In our case, the asymmetry is generated initially by the decay of a singlet $N_3$, therefore not affected by such effects.
 Instead, the lower bound on $N_D$ mass is related to the strength $N_D$ interacts with the singlet and gauge bosons: for a fixed $M_3$ in the TeV range,
 the lower the $m_{N}$, the stronger the  $\Delta N_{D}=1$ washout processes, which tend to maintain $N_3$ in equilibrium.
 The dependence of $Y_{B}$ on the $N_3$ mass scale is manifest in Fig.~\ref{figGammaRate}: a degenerate spectrum, with $M_3$ close to the threshold $m_N+\mu_S$ is favored.~\footnote{Indeed, the triplets $S$ and $N_{D}$ receive thermal mass corrections $\propto g T$, and $N_3$ decays are kinematically open at a temperature $T_D$ actually smaller than $M_3$. The closest $M_3$ is to the $T=0$ threshold $m_N+\mu_S$, the lowest is $T_D$, resulting in a Boltzmann suppression of the washout scatterings involving $N_3$.}
 An absolute lower bound on $m_N$ could be formally derived, however subject to a certain tuning of the parameters.
 Typically $m_N \gtrsim 1.5$ TeV allows our mechanism to work. Such heavy masses cannot be probed at LHC, or oppositely, if a triplet fermion of mass $\lesssim 1$ TeV is observed, the scenario discussed here is not responsible for the observed BAU.
 
 The dependence of the baryon asymmetry on $\mupp$ and $\delta_N$ is presented in the right panel of Fig.~\ref{figGammaRate}.
 For illustration, we fix this time $m_N=\mu_S= 2$ TeV, while $M_3=4.4$ TeV (10\% degeneracy).
 We see the strong influence of $\delta_N$ on $Y_B$ from the rise of the washout effects, implying $\delta_N \lesssim 10^{-6}$. 
 In this case, $\epsilon_{\rm CP}$ increases with $\mupp$ (eq.~\eqref{ApprECP}), as well as the baryon asymmetry. 
 Not illustrated in this plot is the closing of the successful region in green for values of $\mupp$ larger than 3 TeV, due to an increase of the washout processes. However, note that for a given $\mu_S$, $\mupp$ cannot be arbitrarily large, as it enters in DM mass expression \eqref{DMmass}.

 As a concluding remark about this section, we emphasize here the possibility of a low-energy realization of the leptogenesis scenario, although with some tuning of the parameters. Going to higher mass scale for $N_3$ relaxes such tuning.

\section{Discussion and Conclusions}

In this work we consider the model of~\cite{Previous}, implemented with triplet fermions instead of singlets.
The Standard Model particle content is extended with additional fermion and scalar representations:\vspace{-2mm}
\begin{itemize}
\item[\it i)] an extra complex Higgs doublet $H_{2}$;\vspace{-2mm}
\item[\it ii)] a singlet complex scalar $\phi$;\vspace{-2mm}
\item[\it iii)] one ${\rm SU(2)_{\rm L}}$ triplet Dirac fermion $N_{D}$;\vspace{-2mm}
\item[\it iv)] a complex scalar triplet of ${\rm SU(2)_{\rm L}}$  $S$;\vspace{-2mm}
\item[\it v)] a Majorana singlet $N_3$.
\end{itemize}
The particles listed in $i)-iii)$ naturally realize a low scale inverse see-saw mechanism of neutrino mass generation.
 The scalar $iv)$ provides a viable dark matter candidate, and with the two representations in $iii)-v)$ allows us to explain the observed baryon asymmetry of the universe,
 through a \textit{variant} of the standard thermal leptogenesis mechanism.
The overall model is symmetric under a global ${\rm U(1)_{\rm X}}$ transformation, which is spontaneously broken by $H_2$ and $\phi$
vevs at the electroweak scale. The corresponding vevs break the generalized  $B-L$ number (see Table~\ref{FieldAssign}), implying nonzero neutrino masses. Furthermore, this symmetry is broken down to a remnant $Z_{2}$,  which stabilizes the DM. A Goldstone boson emerges from the spontaneous breaking of the symmetry, the Majoron $J$. 

This model is characterized by 6 $Z_{2}$-even scalar particles: 1 charged  $H^{\pm}$, 
3 CP-even neutral scalars $h^{0}$, $H^{0}$ and $h_{A}$ and 2 CP-odd scalar particles $A^{0}$ and $J$.  
The lightest CP-even scalar $h^{0}$ corresponds  to ATLAS~\cite{ATLAS} and CMS~\cite{CMS} observations of a $m_{h^{0}}\simeq 126$ GeV scalar boson. The features of this model may allow us to discriminate it against the SM at  LHC.
 In particular, the presence of a charged scalar  $H^{\pm}$ can reduce or largely enhance  the $h^{0}\to \gamma \gamma$ decay width compared to the SM expectation. Similarly, the invisible decay channel $h^{0}\to J J$ can be sizable and affect the branching ratios of the Higgs boson to the SM particles. Observations of large deviations from the SM case, as well as of other neutral scalar excitations, would constitute the most distinctive signatures of the scalar spectrum presented here. Additional experimental results could help distinguish this model from 2HDM, or other scalar extensions of the SM.

The remnant $Z_{2}$ symmetry guarantees the stability of the DM. We have a total of 4 $Z_{2}$-odd scalars, which arise from
the complex triplet $S$: 2 charged  $S_{L,H}^{\pm}$ and 2 neutral particles  $S_{L,H}^{0}$.
The lightest neutral scalar $S_{L}^{\,0}$  provides a viable DM candidate, as long as the  quartic coupling $\Fs_1^\prime$ contribution to its mass, which splits $m_{S_{L}}^{\,0}$ and $m_{S_{L}}^{\pm}$, is smaller than the one-loop gauge corrections to $m_{S_{L}}^{\pm}$. 
We obtain  numerically  that the lightest viable DM candidate has $m_{\rm DM}\;\gtrsim\; 1290$ GeV, consistently with previous works on scalar triplet DM~\cite{MinimalDM,Hambye:2009pw}.

Light neutrino masses originate from the spontaneous breaking of the generalized $B-L$ charge, instead of relying on explicit lepton number breaking. Two heavy Majorana particles, $\mathcal{N}_{1,2}$ quasidegenerate, are obtained, in addition to two massive and one massless light neutrino.
 Heavy fermions with mass $m_N\lesssim \mathcal{O}(\TeV)$ can in principle be probed at the LHC, thanks to their couplings to SU$(2)$ gauge bosons.

The baryon asymmetry of the universe can be explained within a two-step leptogenesis mechanism~\cite{Previous}, provided one Majorana fermion, $N_{3}$, is included with a direct coupling to both triplet representations $S$ and $N_D$, see eq.~(\ref{neutrL}). 
We obtain that an approximate lower bound on the see-saw scale $m_{N}\gtrsim 1500$ GeV is required to comply with the observed baryon asymmetry. 

The model we present can easily explain and overcome several issues of the standard theory: dark matter abundance, baryon asymmetry of the universe and neutrino oscillation data.  The observation at LHC of a $\mathcal{O}(\TeV)$ fermion triplet, even if  consistent with $N_D$ properties, would exclude the baryogenesis mechanism proposed here, although not the inverse see-saw model itself.
When confronted with additional experimental data, a systematic study of the phenomenology of the model may allow us to significantly constraint the parameter space and eventually rule out this entire scenario.

%%%%%%%%%%%%%%%%%%%%%%%%%%%%%%%%%%%
\section*{Acknowledgements}
We are grateful to D.~Aristizabal Sierra, N.~Bernal, and S.~Palomares-Ruiz for enlightening discussions.
This work is supported by
Funda\c{c}\~{a}o para a Ci\^{e}ncia e a
Tecnologia (FCT, Portugal) through the projects
PTDC/FIS/098188/2008,   CERN/FP/123580/2011
and CFTP-FCT Unit 777,
which are partially funded through POCTI (FEDER).

\end{document}